\documentclass[12pt,twoside,fleqn]{article}
\usepackage{epsfig}
\usepackage[dvips]{rotating}
\usepackage{multirow}
\usepackage{exscale}
\usepackage{amsbsy}
\usepackage{amssymb}
\usepackage[dvips]{color}
\def\lsim{\raise0.3ex\hbox{$<$\kern-0.75em\raise-1.1ex\hbox{$\sim$}}}
\def\gsim{\raise0.3ex\hbox{$>$\kern-0.75em\raise-1.1ex\hbox{$\sim$}}}

\newcommand{\bq}{\begin{equation}}
\newcommand{\eq}{\end{equation}}
\newcommand{\bqa}{\begin{eqnarray}}
\newcommand{\eqa}{\end{eqnarray}}
\newcommand{\bqas}{\begin{eqnarray*}}
\newcommand{\eqas}{\end{eqnarray*}}
\newcommand{\bdm}{\begin{displaymath}}
\newcommand{\edm}{\end{displaymath}}

\newcommand{\Nt}{N_{\tau}}

\newcommand{\nn}{\nonumber}
\begin{document}
\thispagestyle{empty}
\mbox{} \hfill BI-TP 2003/11\\[0mm]
\mbox{} \hfill SWAT/03/374\\
\begin{center}
{\Huge \bf The Equation of State for \\Two Flavor QCD \\at
Non-zero Chemical Potential}

\vspace*{1.0cm}
{\large \bf C.R.~Allton\rlap,$^{\rm a}$ S.~Ejiri\rlap,$^{\rm a,b}$
S.J.~Hands\rlap,$^{\rm a}$ O.~Kaczmarek\rlap,$^{\rm b}$
F.~Karsch\rlap,$^{\rm b}$ 
E.~Laermann$^{\rm b}$ and C.~Schmidt$^{\rm b}$}\\
\vspace*{1.0cm}
{\normalsize
$^{\rm a}$ {Department of Physics, University of Wales Swansea,
          Singleton Park,\\ Swansea SA2 8PP, U.K.} \\
$^{\rm b}$ {Fakult\"at f\"ur Physik, 
Universit\"at Bielefeld, D-33615 Bielefeld, Germany.}
}
\end{center}
\vspace*{1.0cm}
\centerline{\large ABSTRACT}
\baselineskip 20pt
\noindent
We present results of a simulation of QCD on a $16^3\times4$
lattice with 2 continuum flavors of
p4-improved staggered fermion with mass $m/T=0.4$. Derivatives of the
thermodynamic grand potential $\Omega$ with respect to quark chemical
potential $\mu_q$ up to fourth order are calculated, enabling estimates of 
the pressure, quark number density and associated susceptibilities as functions
of $\mu_q$ via 
Taylor series expansion. Discretisation effects associated with various
staggered fermion formulations are discussed in some detail.
In addition it is possible to estimate the radius of
convergence of the expansion as a function of temperature. We also 
discuss the calculation of energy and entropy
densities which are defined via mixed derivatives of $\Omega$ with respect
to the bare couplings and quark masses.

\vfill
\noindent
\mbox{}May 2003\\
\eject
\baselineskip 15pt
\section{Introduction}
Non-perturbative studies of QCD thermodynamics with small but non-zero baryon
charge density by numerical simulation of lattice gauge theory have recently
made encouraging progress \cite{QCDTARO} - \cite{dFP}. 
In particular, it has proved possible to trace out
the pseudocritical line $T_c(\mu_q)$ separating hadronic and quark-gluon plasma
(QGP) phases in the $(\mu_q,T)$ plane out to $\mu_q\simeq O(100)$MeV
\cite{FK} - \cite{dFP}, 
where the quark 
chemical potential $\mu_q$ is the appropriate thermodynamic control variable
in the description of systems with varying particle number using the grand
canonical ensemble. In addition, the first estimate has been made
of the location along this 
line of the critical endpoint expected for $N_{\rm f}=2$ light quark flavors,
where the crossover between hadron and QGP phases becomes a true first order
phase transition 
\cite{FK}. 
As well as being of intrinsic theoretical interest, 
such studies are directly applicable to the regime under current experimental
investigation at RHIC, where corrections to quantities evaluated at $\mu_q=0$ 
are both small and calculable. In this respect it is worth reminding the reader
that in a relativistic heavy ion collision of duration $\sim10^{-22}$s, thermal
equilibration is possible only for processes mediated 
by the strong interaction,
rather than the full electroweak equilibrium achievable, say, in the core of a
neutron star. This means that each quark flavor is a conserved charge,
and conditions at RHIC are thus approximately described by
\begin{equation}
\mu_{\rm u}=\mu_{\rm d}=\mu_q;\;\;\;
\mu_{\rm I}\equiv2(\mu_{\rm u}-\mu_{\rm d})=0;\;\;\;
\mu_{\rm s}=0,
\end{equation}
with $\mu_q\simeq15$MeV \cite{B-MMRS} when we relate the quark and
baryon number chemical potentials via $\mu_B=3\mu_q$. 
In this paper we will present numerical results
for the equation of state, {\it i.e.}  
pressure $p(\mu_q,T)$ and quark number density
$n_q(\mu_q,T)$, obtained  from a lattice QCD simulation with $N_{\rm f}=2$,
which should give a qualitatively correct description of RHIC physics, and
provide a useful warm-up exercise
for the physical case of 2+1 flavors with realistic
light and strange quark masses. 

Direct simulation
using standard Monte Carlo importance sampling is hampered because
the QCD path integral measure $\mbox{det}^{N_{\rm f}}M$, where $M(\mu_q)$
is the Euclidean space fermion kinetic operator, is complex once $\mu_q\not=0$.
In the studies which have appeared to date, two
fundamentally distinct approaches to this problem have emerged. 
In the {\em reweighting} method results from simulations at $\mu_q=0$
are reweighted on a configuration-by-configuration basis with the
correction factor $[\mbox{det}M(\mu_q)/\mbox{det}M(0)]^{N_{\rm f}}$ yielding
formally exact estimates for expectation values. Indeed, it is found that
if reweighting is performed simultaneously in two or more parameters,
convergence of this method on 
moderately-sized systems is considerably enhanced \cite{FK2}. 
This method has been used 
on lattice sizes up to $12^3\times4$ with $N_{\rm f}=2+1$ 
to map out the pseudocritical line and estimate the location of
the critical endpoint at $\mu_q^{crit}\simeq240$MeV, $T^{crit}\simeq160$MeV
\cite{FK}.
More recently the equation of state in the entire region to the left of the 
endpoint has been calculated this way \cite{FKS}. 
However, it remains unclear whether
the thermodynamic limit can be reached using this technique.

{\em Analytic} approaches use data from
regions where direct simulation is possible, either by calculating derivatives
with respect to $\mu_q$ (or more properly with respect
to the dimensionless combination $\mu_q/T$) to construct a Taylor expansion
for quantities of interest \cite{QCDTARO,us,Gottlieb,GG},
or more radically by analytically continuing results from simulations with
imaginary $\mu_q$ (for which the integration measure remains real) 
to real $\mu_q$. The second technique has been used to map $T_c(\mu_q)$
for QCD with both $N_{\rm f}=2$ \cite{dFP} and $N_{\rm f}=4$ \cite{dEL}, 
in the latter case 
finding evidence that the line is first order in nature.
Fortunately, the pseudocritical line found in \cite{dFP}
is in reasonable agreement with that found by reweighting; moreover 
the radius of convergence within which analytic continuation from imaginary
$\mu_q$
is valid corresponds to $\mu_q/T\leq\pi/3$ \cite{Laine}. The leading 
non-trivial term of quadratic order in the Taylor expansion 
appears to provide a good approximation
throughout this region. In general though, while analytic approaches have no
problem approaching the thermodynamic limit, it is not yet clear if and how 
they can be extended into the region around the critical endpoint 
(but see \cite{Christian}), and to observables that vary strongly with 
$\mu_q$ like eg. the pressure or energy density.

In our previous paper \cite{us} we used a hybrid of the two techniques, by 
making a Taylor series estimate of the reweighting factor 
$[\mbox{det}M(\mu_q)/\mbox{det}M(0)]^{N_{\rm f}}$ to $O(\mu_q^2)$. 
Since this is
considerably cheaper than a calculation of the full determinant, we are able to
explore a larger $16^3\times4$ system, and also exploit an
improved action in both gauge and fermion \cite{Heller} sectors, 
thus dramatically reducing
discretisation artifacts on what at $T_c(\mu_q=0)\simeq170$MeV 
is still a coarse lattice.
Our results yield a curvature of the phase transition line
$T_c(d^2T_c/d\mu_q^2)\vert_{\mu_q=0}\simeq-0.14(6)$, consistent with the other
approaches. Although our simulation employs quark masses $m/T=0.4,0.8$, 
not yet realistically light, the results also suggest that
any dependence on quark mass is weak. In the current paper we extend the Taylor
series to the next order $O(\mu_q^4)$ but this time remain entirely
within the analytic framework, using derivatives calculated at $\mu_q=0$ to 
evaluate non-zero density corrections to the pressure $p$ and quark number
susceptibility $\chi_q\equiv\partial n_q/\partial\mu_q$, as well as the quark
number density $n_q$ itself. In fact, since the correction $\Delta p$ 
can be evaluated at fixed temperature, it turns out to be considerably 
easier to calculate than the equation of state at $\mu_q=0$
\cite{KLP,CPPACS}. Since we now have the first two non-trivial terms in the
Taylor expansion, we are also able to estimate its radius of convergence
as a function of $T$, 
and confirm that close to $T_c(\mu_q=0)$ 
the results of our previous study for the
critical line curvature can be trusted out to $O(100\mbox{MeV})$, whereas
at higher temperatures a considerably larger radius of convergence is likely to
be found.
Finally we consider mixed derivatives 
with respect to both $\mu_q$ and the other
bare parameters $\beta$ and $m$, which are required to estimate energy
$\epsilon$ and entropy $s$ densities. Due to the presence of a critical
singularity, these latter quantities appear considerably harder to calculate in
the critical region using this approach.

Sec.~\ref{sec:form} outlines the formalism used in the calculation and 
specifies
which derivatives are required. In Sec.~\ref{cut-off_dependence} we present
a calculation of the cutoff dependence of the terms in the expansion 
for standard, and for Naik- and p4-improved staggered lattice fermions, 
showing that
both improvements result in a dramatic reduction of discretisation effects.
Our numerical results are presented in Sec.~\ref{sec:results}, and a brief
discussion in Sec.~\ref{sec:andfinally}. Two appendices contain further 
details on the calculation of the required derivatives and the cutoff
dependence.

\section{Formulation}
\label{sec:form}

In the grand canonical ensemble pressure is given in terms of the grand
partition function ${\cal Z}(T,V,\mu_q)$ by
\begin{equation}
{p\over T^4}={1\over{VT^3}}\ln{\cal Z}.
\end{equation}
Note that we have been careful to express both sides of this relation in
dimensionless quantities. Since the free energy and its derivatives can only be
calculated using conventional Monte Carlo methods at $\mu_q=0$, 
we proceed by making a Taylor
expansion about this point in powers of the dimensionless quantity $\mu_q/T$:
\begin{eqnarray}
\hspace{-0.2cm}\Delta\left({p\over T^4}(\mu_q)\right)\equiv
{p\over T^4}\biggr\vert_{T,\mu_q}-{p\over T^4}\biggr\vert_{T,0} \hspace{-0.2cm}
&=&
{1\over2!}{\mu_q^2\over T^2}
{{\partial^2(p/T^4)}\over{\partial(\mu_q/T)^2}}+
{1\over4!}
{\mu_q^4\over T^4}
{{\partial^4(p/T^4)}\over{\partial(\mu_q/T)^4}}
+\cdots \nonumber \\
&\equiv& \sum_{p=1}^{\infty} c_p(T) \biggl( {\mu_q \over T}\biggr)^p
\label{eq:taylorcont}
\end{eqnarray}
where derivatives are taken at $\mu_q=0$.
Note that calculating $\Delta(p/T^4)$ is considerably easier than 
$p(T,\mu_q=0)$ itself, because whereas $\ln{\cal Z}$ must be estimated by 
integrating along a trajectory in the bare parameter plane
\cite{KLP,CPPACS}, 
its derivatives can be related to observables which are 
directly simulable at fixed $(\beta,m)$, where $\beta$ is the gauge coupling 
parameter and $m$ the bare quark mass.
Only even powers appear in (\ref{eq:taylorcont})
because as shown in \cite{us}, odd derivatives of the 
free energy with respect to $\mu_q$ vanish at this point.
Note also that we will work throughout with fixed bare mass, implying that 
our computation of $\Delta(p/T^4)$ is strictly valid along a line of fixed
$m/T$.

For QCD with staggered fermions the partition function may be written
\begin{equation}
{\cal Z}=\int{\cal D}U(\mbox{det} M)^{N_{\rm f}/4}\exp(-S_g),
\end{equation}
with $U\in SU(3)$ denoting the gauge field variables,
$S_g[U]$ the link action and $M[U;\mu_q]$ the kinetic operator describing
a single staggered fermion, equivalent to $N_{\rm f}=4$ continuum flavors.
On a lattice of size $N_\sigma^3\times N_\tau$ 
with physical lattice spacing $a$, so
that $T=(aN_\tau)^{-1}$, we define a dimensionless
lattice chemical potential variable
$\mu=\mu_q a$. 
Equation (\ref{eq:taylorcont}) then becomes
\begin{equation}
\Delta\left({p\over T^4}\right)=
{1\over2}{N_\tau^3\over N_\sigma^3}
\mu^2{{\partial^2\ln{\cal Z}}\over{\partial\mu^2}}+
{1\over24}{N_\tau^3\over N_\sigma^3}\mu^4
{{\partial^4\ln{\cal Z}}\over{\partial\mu^4}}+
\cdots
\label{eq:deltap}
\end{equation}
The derivatives may be expressed as expectation values 
evaluated at $\mu=0$:
\begin{equation}
\frac{\partial^2 \ln {\cal Z}}{\partial \mu^2} =
\left\langle \frac{N_{\rm f}}{4}
\frac{\partial^2 (\ln \det M)}{\partial \mu^2} \right\rangle
+\left\langle \left( \frac{N_{\rm f}}{4}
\frac{\partial (\ln \det M)}{\partial \mu} \right)^2 \right\rangle,
\label{eq:d2lnz}
\end{equation}
\vspace{-0.4cm}
\begin{eqnarray}
&& \hspace{-1.6cm}\frac{\partial^4 \ln {\cal Z}}{\partial \mu^4} 
\phantom{E=}\hspace{-0.3cm}=
\left\langle \frac{N_{\rm f}}{4}
\frac{\partial^4 (\ln \det M)}{\partial \mu^4} \right\rangle
+ 4 \left\langle \left( \frac{N_{\rm f}}{4} \right)^2
\frac{\partial^3 (\ln \det M)}{\partial \mu^3}
\frac{\partial (\ln \det M)}{\partial \mu} \right\rangle 
\label{eq:d4lnz}\\
&&\hspace{-1.6cm} + 3 \left\langle \left( \frac{N_{\rm f}}{4} \right)^2
\left( \frac{\partial^2 (\ln \det M)}{\partial \mu^2}
\right)^2 \right\rangle
+ 6 \left\langle \left( \frac{N_{\rm f}}{4} \right)^3
\frac{\partial^2 (\ln \det M)}{\partial \mu^2}
\left( \frac{\partial (\ln \det M)}{\partial \mu} \right)^2
\right\rangle \nonumber \\
&&\hspace{-1.6cm} + \left\langle \left( \frac{N_{\rm f}}{4}
\frac{\partial (\ln \det M)}{\partial \mu} \right)^4 \right\rangle
- 3 \left[ \left\langle \frac{N_{\rm f}}{4}
\frac{\partial^2 (\ln \det M)}{\partial \mu^2} \right\rangle
+ \biggl\langle \left( \frac{N_{\rm f}}{4}
\frac{\partial (\ln \det M)}{\partial \mu} \right)^2 \biggr\rangle
\right]^2 . \nonumber
\end{eqnarray}
All expectation values are calculated using the measure
${\cal Z}^{-1}(\mu=0){\cal D}U(\mbox{det}M[\mu=0])^{N_{\rm f}/4}e^{-S_g}$ and
in deriving (\ref{eq:d2lnz}) and (\ref{eq:d4lnz}) we used the fact that 
$\langle\partial^n(\ln\mbox{det}M)/\partial\mu^n\rangle=0$ for $n$ odd.
To evaluate these expressions we need the following explicit forms:
\begin{eqnarray}
\label{eq:dermu1}
&&\hspace{-1.2cm}\frac{\partial (\ln \det M)}{\partial \mu}\hspace{1.2cm}
= {\rm tr} \left( M^{-1} \frac{\partial M}{\partial \mu} \right) \\
\label{eq:dermu2}
&&\hspace{-1.2cm}\frac{\partial^2 (\ln \det M)}{\partial \mu^2}
\hspace{1.2cm}
= {\rm tr} \left( M^{-1} \frac{\partial^2 M}{\partial \mu^2} \right)
 - {\rm tr} \left( M^{-1} \frac{\partial M}{\partial \mu}
                   M^{-1} \frac{\partial M}{\partial \mu} \right) \\
\label{eq:dermu3}
&&\hspace{-1.2cm}\frac{\partial^3 (\ln \det M)}{\partial \mu^3}
\hspace{1.2cm}
= {\rm tr} \left( M^{-1} \frac{\partial^3 M}{\partial \mu^3} \right)
 -3 {\rm tr} \left( M^{-1} \frac{\partial M}{\partial \mu}
              M^{-1} \frac{\partial^2 M}{\partial \mu^2} \right) \\
&& \hspace{2.8cm}+2 {\rm tr} \left( M^{-1} \frac{\partial M}{\partial \mu}
        M^{-1} \frac{\partial M}{\partial \mu}
        M^{-1} \frac{\partial M}{\partial \mu} \right) \nonumber \\
\label{eq:dermu4}
&&\hspace{-1.2cm}\frac{\partial^4 (\ln \det M)}{\partial \mu^4}
\hspace{1.2cm}
= {\rm tr} \left( M^{-1} \frac{\partial^4 M}{\partial \mu^4} \right)
 -4 {\rm tr} \left( M^{-1} \frac{\partial M}{\partial \mu}
              M^{-1} \frac{\partial^3 M}{\partial \mu^3} \right) \\
&& -3 {\rm tr} \left( M^{-1} \frac{\partial^2 M}{\partial \mu^2}
        M^{-1} \frac{\partial^2 M}{\partial \mu^2} \right)
 +12 {\rm tr} \left( M^{-1} \frac{\partial M}{\partial \mu}
        M^{-1} \frac{\partial M}{\partial \mu}
        M^{-1} \frac{\partial^2 M}{\partial \mu^2} \right) \nonumber \\
&& -6 {\rm tr} \left( M^{-1} \frac{\partial M}{\partial \mu}
        M^{-1} \frac{\partial M}{\partial \mu}
        M^{-1} \frac{\partial M}{\partial \mu}
        M^{-1} \frac{\partial M}{\partial \mu} \right) \nonumber
\end{eqnarray}
The traces can be estimated using the stochastic method reviewed in \cite{us}.
Since $\partial^nM/\partial\mu^n$ is a local operator, expressions containing
$p$ powers of $M^{-1}$ require $p$ operations of matrix inversion on a vector.

Next we discuss the quark number density $n_q$ and its fluctuations.
Starting from the Maxwell relation
\begin{equation}
n_q=-{{\partial^2\Omega}\over{\partial V\partial\mu_q}}=
{{\partial N_q}\over{\partial V}}={{\partial p}\over{\partial\mu_q}},
\end{equation}
where $\Omega=-T\ln{\cal Z}$ is the thermodynamic grand potential
and $N_q$ the net quark number,
we can write an equation for the quark number density $n_q$ analogous to
(\ref{eq:deltap}):
\begin{equation}
{n_q\over T^3}=
{N_\tau^2\over N_\sigma^3}
\mu{{\partial^2\ln{\cal Z}}\over{\partial\mu^2}}+
{1\over6}{N_\tau^2\over N_\sigma^3}\mu^3
{{\partial^4\ln{\cal Z}}\over{\partial\mu^4}}+
\cdots
\label{eq:density}
\end{equation}
It is also possible to interpret 
derivatives of $p$ with respect to $\mu_q$ in terms
of the various susceptibilities giving information on number density
fluctuations \cite{Gottlieb,Liu}. We define quark number $(q)$, 
isospin $(I)$ and
charge $(C)$ susceptibilities as follows:
\begin{eqnarray}
\frac{\chi_q}{T^2}
&=& \left( \frac{\partial}{\partial (\mu_{\rm u}/T)}
+\frac{\partial}{\partial (\mu_{\rm d}/T)} \right)
{{n_{\rm u} + n_{\rm d}}\over T^3}, \label{eq:chisq}\\
\frac{\chi_{\rm I}}{T^2}
&=& {1\over4}\left( \frac{\partial}{\partial (\mu_{\rm u}/T)}
-\frac{\partial}{\partial (\mu_{\rm d}/T)} \right)
{{n_{\rm u} - n_{\rm d}}\over T^3}, \label{eq:chist}\\
\frac{\chi_{\rm C}}{T^2}
&=& \left( \frac{2}{3} \frac{\partial}{\partial (\mu_{\rm u}/T)}
-\frac{1}{3} \frac{\partial}{\partial (\mu_{\rm d}/T)} \right)
{{2n_{\rm u}-n_{\rm d}}\over{3T^3}}\label{eq:chic}.
\end{eqnarray}
Quark and baryon number susceptibilities are related 
by $\chi_{\rm B}\equiv\partial n_{\rm B}/\partial\mu_{\rm B}=3^{-2}\chi_q$.
Any difference between  $\chi_q$ and $4\chi_{\rm I}$ is due to correlated
fluctuations in the individual densities of $u$ and $d$ quarks.
With the choice $\mu_{\rm u}=\mu_{\rm d}=\mu_q=\mu a^{-1}$, $m_{\rm u}=m_{\rm
d}$,
which approximates the physical 
conditions at RHIC, $\chi_q$ can then be expanded in terms of
quantities already used in the calculation of $p$ and $n_q$:
\begin{equation}
{\chi_q\over T^2}\biggr\vert_{\mu_q=0}
\equiv{1\over T^2}{{\partial n_q}\over{\partial\mu_q}}
={N_\tau\over N_\sigma^3}{{\partial^2\ln{\cal Z}}\over\partial\mu^2};\;\;\;
{{\partial^2(\chi_q/T^2)}\over\partial(\mu_q/T)^2}\biggr\vert_{\mu_q=0}
={1\over{N_\tau N_\sigma^3}}
{{\partial^4\ln{\cal Z}}\over{\partial\mu^4}},
\label{eq:susc}
\end{equation}
whereas the expansion of $\chi_{\rm I}$ 
is determined by the following expectation values:
\begin{eqnarray}
&&\hspace{-1.4cm}\frac{\chi_{\rm I}}{T^2}\biggr\vert_{\mu_q=0}
= \frac{N_\tau}{4N_\sigma^3} \left\langle \frac{2}{4}
\frac{\partial^2 (\ln \det M)}{\partial \mu^2} \right\rangle ,
\\
&&\hspace{-2.0cm}
\frac{\partial^2 (\chi_{\rm I}/T^2)}{\partial
(\mu_{q}/T)^2}\biggr\vert_{\mu_q=0}
=
\frac{1}{4N_\sigma^3 N_\tau} \left[
\left\langle \frac{2}{4}
\frac{\partial^4 (\ln \det M)}{\partial \mu^4} \right\rangle
+ 2 \left\langle \left( \frac{2}{4} \right)^2
\frac{\partial^3 (\ln \det M)}{\partial \mu^3}
\frac{\partial (\ln \det M)}{\partial \mu} \right\rangle \right.\nonumber \\
&&\hspace{-1.0cm}+\left\langle \left( \frac{2}{4} \right)^2
\left( \frac{\partial^2 (\ln \det M)}{\partial \mu^2}
\right)^2 \right\rangle
+ \left\langle \left( \frac{2}{4} \right)^3
\frac{\partial^2 (\ln \det M)}{\partial \mu^2}
\left( \frac{\partial (\ln \det M)}{\partial \mu} \right)^2
\right\rangle \nonumber \\
&& \hspace{-1.0cm}-\left[ \left\langle \frac{2}{4}
\frac{\partial^2 (\ln \det M)}{\partial \mu^2} \right\rangle
+ \left. \left\langle \left( \frac{2}{4}
\frac{\partial (\ln \det M)}{\partial \mu} \right)^2 \right\rangle
\right] \left\langle \frac{2}{4}
\frac{\partial^2 (\ln \det M)}{\partial \mu^2} \right\rangle
\right] ,\label{eq:chit} 
\end{eqnarray}
where we have explicitly set $N_{\rm f}=2$.
Charge fluctuations are then given by the relation
\begin{equation}
\frac{\chi_{\rm C}}{T^2}
= \frac{1}{36} \frac{\chi_q}{T^2}
+ \frac{\chi_{\rm I}}{T^2}
+ \frac{1}{6}
\left( \frac{\partial(n_{\rm u}/T^3)}{\partial (\mu_{\rm u}/T)}
- \frac{\partial(n_{\rm d}/T^3)}{\partial (\mu_{\rm d}/T)} \right), 
\end{equation}
where the third term vanishes for $\mu_{\rm u}=\mu_{\rm d}$, $m_{\rm u}=m_{\rm
d}$.

Finally we discuss the energy density $\epsilon$, most conveniently 
extracted using the conformal anomaly relation
\begin{equation}
{{\epsilon-3p}\over T^4}
=-{1\over{VT^3}}\left[a{{\partial\beta}\over{\partial a}}
{{\partial\ln{\cal Z}}\over{\partial\beta}}+
a{{\partial m}\over{\partial a}}
{{\partial\ln{\cal Z}}\over{\partial m}}\right],
\label{eq:conformal}
\end{equation}
where $\beta$ and $m$ are the bare coupling and quark mass respectively.
In fact, for $\mu\not=0$ the derivation of this expression needs careful
discussion. Start from the defining relation
\begin{equation}
\Omega=E-T{\cal S}-\mu_qN_q=-pV=-T\ln{\cal Z},
\end{equation}
where ${\cal S}$ is entropy.
For a Euclidean action $S=S(\beta,m,\mu)$ 
defined on an isotropic lattice of spacing $a$
we have the identity
\begin{equation}
a{{dS}\over{da}}=3V{{\partial S}\over{\partial V}}-T{{\partial S}\over{\partial
T}}.
\end{equation}
It follows that
\begin{eqnarray}
V{{\partial\Omega}\over{\partial V}}&=&
VT\left\langle{{\partial S}\over{\partial V}}
\right\rangle=-pV\\
T{{\partial\Omega}\over{\partial T}}&=&
\Omega+T^2\left\langle{{\partial S}\over{\partial T}}\right\rangle
=-T{\cal S}=\Omega-E+\mu_qN_q
\end{eqnarray}
implying
\begin{equation}
\epsilon-3p-\mu_qn_q=
{T\over V}\left\langle a{{\partial S}\over{\partial a}}\right\rangle
=-{T\over V}
\left[a{{\partial\beta}\over{\partial a}}
{{\partial\ln{\cal Z}}\over{\partial\beta}}+
a{{\partial m}\over{\partial a}}
{{\partial\ln{\cal Z}}\over{\partial m}}+
a{{\partial\mu }\over{\partial a}}
{{\partial\ln{\cal Z}}\over{\partial\mu}}\right]
\end{equation}
where we have allowed for the dependence of the lattice action on all
bare parameters. Since however $\mu\equiv\mu_qa$, and a parameter
multiplying a conserved
charge experiences no renormalisation, the third terms on each
side cancel leaving the relation (\ref{eq:conformal}).

Taylor expansion of (\ref{eq:conformal})
about $\mu=0$ leads to the expression
\begin{eqnarray}
\Delta\left({{\epsilon-3p}\over T^4}\right)=
&-&a{{\partial\beta}\over{\partial a}}{N_\tau^3\over N_\sigma^3}
\left[{1\over2}\mu^2{{\partial^3\ln{\cal Z}}\over{\partial\beta\partial\mu^2}}
+{1\over24}\mu^4{{\partial^5\ln{\cal
Z}}\over{\partial\beta\partial\mu^4}}+\cdots\right]\nonumber\\
&-&a{{\partial m}\over{\partial a}}{N_\tau^3\over N_\sigma^3}
\left[{1\over2}\mu^2{{\partial^3\ln{\cal Z}}\over{\partial m\partial\mu^2}}
+{1\over24}\mu^4{{\partial^5\ln{\cal
Z}}\over{\partial m\partial\mu^4}}+\cdots\right].
\label{eq:deltaeps}
\end{eqnarray}
The beta function $a(\partial\beta/\partial a)$ may be estimated by
measurements of observables at $(T,\mu_q)=(0,0)$; 
the factor $a(\partial m/\partial
a)$ is poorly constrained by current lattice data  but vanishes in the
chiral limit, so is frequently neglected. In order to assess the magnitude
of the resulting error, it is nonetheless 
useful to calculate all the derivative terms. 
They may be estimated using the formul\ae
\begin{eqnarray}
\frac{\partial \langle {\cal O} \rangle}{\partial \beta} &=&
\left\langle {\cal O} \left(-\frac{\partial S_g}{\partial \beta} \right)
\right\rangle -\langle {\cal O} \rangle \left\langle
-\frac{\partial S_g}{\partial \beta} \right\rangle;\\
\frac{\partial \langle {\cal O} \rangle}{\partial m} &=&
\left\langle \frac{\partial {\cal O}}{\partial m} \right\rangle
+ \left\langle {\cal O}  \frac{N_{\rm f}}{4}
\frac{\partial (\ln \det M)}{\partial m}
\right\rangle -\langle {\cal O} \rangle \left\langle
\frac{N_{\rm f}}{4} 
\frac{\partial (\ln \det M)}{\partial m} \right\rangle.
\end{eqnarray}
The derivative $\partial S_g/\partial\beta$ is, of course, simply the
combination of plaquettes comprising the gauge action itself, and
derivatives with respect to $m$ can be evaluated using
\begin{equation}
\frac{\partial^{n+1} (\ln \det M)}{\partial m\partial \mu^n}
= \frac{\partial^n ({\rm tr} M^{-1})}{\partial \mu^n}.
\end{equation}
The implementation of the second square bracket in (\ref{eq:deltaeps}) in
terms of lattice operators is straightforward but unwieldy; for reference the
non-vanishing terms are listed in Appendix~\ref{appendix:eps}.

\section{Analyzing the cut-off dependence}
\label{cut-off_dependence}
In this section we discuss the influence of a
non-zero chemical potential $\mu_q$ on the cut-off effects present
in calculations of bulk thermodynamic observables
on a lattice with finite temporal extent $N_\tau$. For $\mu_q=0$
this issue has been discussed
extensively  for both gluonic and fermionic sectors of QCD.
In particular, it has been shown that the use of improved actions
is mandatory if one wants to ensure that discretisation errors in
the calculation of quantities like the pressure $p$ or energy density
$\epsilon$ are below the 10\% level on moderately sized
lattices $N_\tau \lsim (8-10)$ \cite{Heller}. We now want to extend these
considerations to the case $\mu_q\not=0$, which affects the 
quark sector only. Following \cite{Heller} we will
concentrate on an evaluation of the pressure. As we 
will be evaluating thermodynamic quantities
using a Taylor expansion in $\mu_q/T$ we want to understand
the cut-off dependence of $p(\mu_q)$ and its expansion
coefficients in terms of $\mu_q / T$.

In the limit of high temperature or density, due to asymptotic
freedom thermodynamic observables like
$p$ or $\epsilon$ are
expected to approach their free gas, {\it i.e.}  Stefan-Boltzmann (SB) values.
In this limit cut-off effects become most significant as the relevant
momenta of partons contributing to the thermodynamics are $O(T)$
and thus of similar magnitude
to the UV cut-off $a^{-1}$. Short distance properties thus dominate
ideal gas behaviour and cut-off effects are controlled by the lattice
spacing expressed in units of the temperature, $Ta\equiv 1/N_\tau$.

In the continuum the pressure of an ideal gas of quarks and anti-quarks
is given by
\bq
\hspace{-0.4cm}\left.\frac{p}{T^4}\right|_{\infty} = \frac{N_{\rm f}}
{2\pi^2 T^3} \int_0^\infty \!\!\!dk\,k^2
\ln\left[(1+z\exp\{-\varepsilon(k)/T\})
(1+z^{-1}\exp\{-\varepsilon(k)/T\})\right]
\label{pressure_con}
\eq
with the fugacity $z\equiv\exp\{\mu_q/T\}$ and the relativistic single particle
energies $\varepsilon(k)=\sqrt{k^2+m^2}$. For massless quarks
one finds from an evaluation of the integral the pressure
as a finite polynomial in $\mu_q / T$:
\bq
\left.\frac{p}{T^4}\right|_{\infty} = \frac{7N_{\rm f}\pi^2}{60}
+ \frac{N_{\rm f}}{2}\left(\frac{\mu_q}{T}\right)^2
+ \frac{N_{\rm f}}{4\pi^2}
\left(\frac{\mu_q}{T}\right)^4. \label{Stefan-Boltzmann}
\eq
For $m$ non-zero the pressure is a series in the fugacity:
\bq
\frac{p}{T^4} = \left( \frac{m}{T} \right)^2 \frac{N_{\rm f}}{2\pi^2}
\sum_{\ell=1}^{\infty} (-1)^{\ell+1}\,\ell^{-2}\, K_2(\ell m/T)\, 
(z^\ell+z^{-\ell}),
\label{bessel}
\eq
where $K_2$ is a Bessel function. Of course, Eq.~(\ref{bessel}) can
also be reorganised as a power series in $\mu_q / T$.

It is well known that the
straightforward lattice representation of the QCD partition function
in terms of the standard Wilson gauge and 
staggered fermion actions leads to a systematic $O(a^2)$ cut-off
dependence of physical observables. In the infinite temperature 
limit this gives rise
to $O((aT)^2\equiv 1/\Nt^2)$ deviations of the pressure from the
SB value (\ref{Stefan-Boltzmann});
\bq
\left.\frac{p}{T^4}\right|_{\Nt}=\left.\frac{p}{T^4}\right|_{\infty} +
\frac{d}{\Nt^2} + O(\Nt^{-4}).
\label{eq:cutcorr}
\eq
Using improved discretisation schemes it is possible to ensure that cut-off 
effects only start to contribute at $O(N_\tau^{-4})$ \cite{Naik},
or to considerably reduce the magnitude of the coefficient $d$ 
relative to the standard discretisation scheme for staggered fermions
\cite{Heller}.

\begin{figure}[tb]
\begin{center}
\begin{minipage}[c][7.8cm][c]{7.4cm}
\begin{center}
\epsfig{file=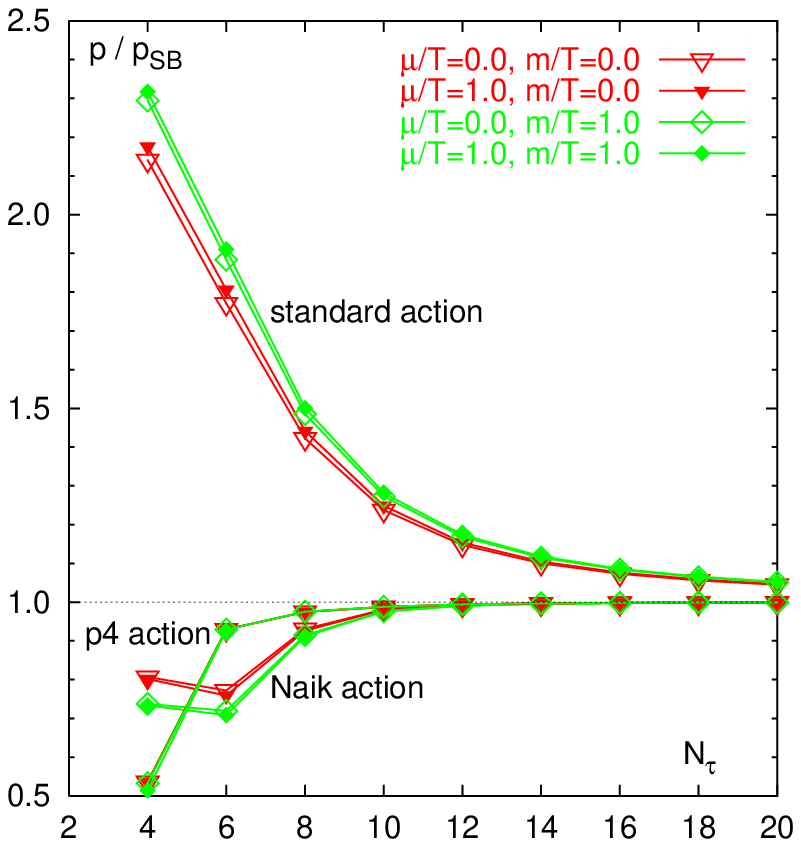, width=7.4cm}\\[-1mm]
(a)
\end{center}
\end{minipage}
\begin{minipage}[c][7.8cm][c]{7.4cm}
\begin{center}
\epsfig{file=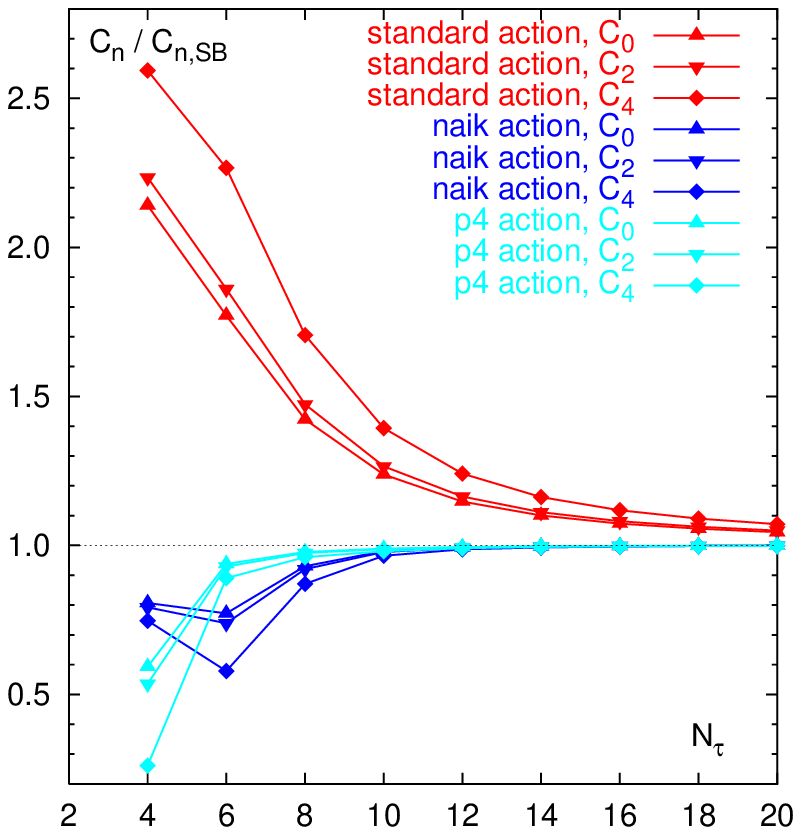, width=7.4cm}\\[-1mm]
(b)
\end{center}
\end{minipage}
\caption{The pressure calculated on lattices with temporal extent
$N_\tau$ in units of the continuum ideal Fermi gas value.
(a) shows results for the standard, Naik and p4
  actions at $(\mu_q/T,\, m/T)=(0,\,0),\,(0,\,1),(1,\,0)$ and $(1,\,1)$;
(b) the coefficients ${\cal
  C}_0,\,{\cal C}_2,\,{\cal C}_4$ of the $\mu_q/T$ expansion of 
  $p(m/T=0)$ divided by the corresponding SB constant as a
  function of $\Nt$.}
\label{fig:pressure_cutoff_mu}
\end{center}
\end{figure}

For $\mu_q=0$ the pressure of free staggered fermions on
lattices with infinite spatial volume ($N_\sigma = \infty$) 
but finite temporal
extent $\Nt$ is given by
\bqa
\left.\frac{p}{T^4}\right|_{\Nt}&=&\frac{3}{8}N_f \Nt^4 \frac{1}{(2\pi)^3}
\int_0^{2\pi}d^3\vec p \Bigg[ \Nt^{-1} \sum_{p_4} \ln\left(\omega^2(p) + 4
  f^2_4(p)\right) \nn \\
&& - \frac{1}{2\pi}\int_0^{2\pi} dp_4 \ln\left(\omega^2(p) + 4
  f^2_4(p)\right)\Bigg]. \label{pressure_lat_pert}
\eqa
In the first term the sum $\sum_{p_4}$ runs over all discrete Matsubara modes,
i.e. $p_4\in\{ (2n+1)\pi/\Nt|n=0,\dots, \Nt-1\}$, whereas in the second
term we have an integral over $p_4$ which gives the vacuum contribution.
For quarks of mass $m$
the function $\omega^2(p)$ is given by
$\omega^2(p)\equiv 4\sum_{\mu=1}^3 f_\mu^2(p) +N_\tau^{-2}(m/T)^2$. Here 
we have introduced functions $f_\mu(p)$ to describe the momentum
dependence of the propagator for the standard, Naik \cite{Naik} and 
p4 staggered fermion actions \cite{Heller}:
\bqa
f_\mu(p) &=& \frac{1}{2} \sin p_\mu \qquad \mbox{(standard staggered action)}
\\
f_\mu(p) &=& \frac{9}{16} \sin p_\mu - \frac{1}{48} \sin3p_\mu \qquad
  \mbox{(Naik action)} \\
f_\mu(p) &=& \frac{3}{8} \sin p_\mu + \frac{2}{48} \sin p_\mu\sum_{\nu\ne\mu}
\cos2p_\nu \qquad \mbox{(p4 action)}.
\label{propp}
\eqa
The introduction of a non-zero chemical potential is easily achieved by
substituting every temporal momentum $p_4$
by $p_4-i\mu\equiv p_4-i\Nt^{-1}(\mu_q/T)$. The integrals in
(\ref{pressure_lat_pert}) have been evaluated numerically for
different $N_\tau$. Results for different
values of $\mu_q /T$ and $m/T$ are shown for the different fermion actions in
Fig.~\ref{fig:pressure_cutoff_mu}.
For the standard action cutoff effects remain $\geq10$\% out to 
$N_\tau\approx16$, whereas both improved actions are hard to distinguish from
the continuum result by $N_\tau=10$.
We note that
lines for different $\mu_q/T$ values but the same quark mass 
fall almost on top of each other. Cutoff effects are thus 
almost independent of $\mu_q$. The effect of $\mu_q\not=0$
on the cutoff
dependence of the pressure is even smaller than the effect of quark mass
$m\not=0$.

As can be seen from (\ref{Stefan-Boltzmann}) for moderate values
of $\mu_q/T$ the $\mu$-dependence of the continuum ideal gas pressure is
dominated by the leading $O((\mu_q /T)^2)$ contribution. In order
to control the cut-off dependence of the various expansion terms
we have expanded the
integrand of (\ref{pressure_lat_pert}) up to order 
$O((\mu_q/T)^6)$. For the standard action the series starts with
\bqa
\lefteqn{\ln\left(\omega^2(p)+\sin^2\left(p_4-\frac{i}{\Nt}\frac{\mu_q}{
        T}\right)\right) \,=\, \ln D} \nn \\
&-& \frac{2i\cos p_4\sin p_4}{D\,\Nt}\left(\frac{\mu_q}{T}\right) \nn \\
&-& \frac{-1 + 4\,D\,\cos 2p_4 + \cos 4p_4}
        {4\,D^2\,\Nt^2}\left(\frac{\mu_q}{T}\right)^2 \nn \\
&-& \frac{i \,\left( -1 + 4\,D^2 +
      6\,D\,\cos 2p_4 + \cos 4p_4 \right) \,
    \sin 2p_4}{6\,D^3\,\Nt^3}\left(\frac{\mu_q}{T}\right)^3 \nn \\
&+& O\left(\left(\frac{\mu_q}{T}\right)^4\right).
\label{pressure_lat_series}
\eqa
Here we use the shorthand
notation
$D=4\sum_{\mu=1}^4 f_\mu^2(p)$. 
The remaining orders as well as the series for Naik and p4 actions are
given in Appendix~\ref{pressure_mu_series}. 
A common feature of these expansions is
that the odd terms are pure imaginary and the integral and sum over
$p_4$ of those terms vanish due to a factor $\sin(np_4)$ which always
appears. To be more precise, this factor always forms the pattern
$\sin(np_4)\cos(mp_4)$ which can be shown to vanish,
 either after summation over the 
discrete set of $p_4$ values, or integration
from $0$ to $2\pi$, for $n,\,m\in \mathbb{N}$.
Performing the momentum integration and the summation over Matsubara
modes we obtain
the coefficients of the ${\mu_q/T}$ expansion of the  pressure;
\bq
\frac{p}{T^4}\Big|_{\Nt} = N_{\rm f}\sum_{i=0}^{\infty} {\cal C}_i\Big|_{\Nt}
 \left(\frac{\mu_q}{T}\right)^i \quad .
\label{eq:defcoeff}
\eq
We checked numerically that with increasing $\Nt$ the coefficients
${\cal C}_0$, ${\cal C}_2$ and
 ${\cal C}_4$ do indeed converge to their
corresponding SB values,
\begin{equation}
\lim_{\Nt\to\infty} {\cal C}_0 = \frac{7\pi^2}{60}; \quad 
\lim_{\Nt\to\infty} {\cal C}_2 = \frac{1}{2}; \quad  
\lim_{\Nt\to\infty} {\cal C}_4 = \frac{1}{4\pi^2}.
\end{equation}
Fig.~\ref{fig:pressure_cutoff_mu}b shows ${\cal C}_0$, ${\cal C}_2$ and
 ${\cal C}_4$
for the standard, Naik and p4 actions with massless quarks, normalized
to the corresponding SB value.
\begin{figure}[tb]
\begin{center}
\begin{minipage}[c][7.8cm][c]{7.4cm}
\begin{center}
\epsfig{file=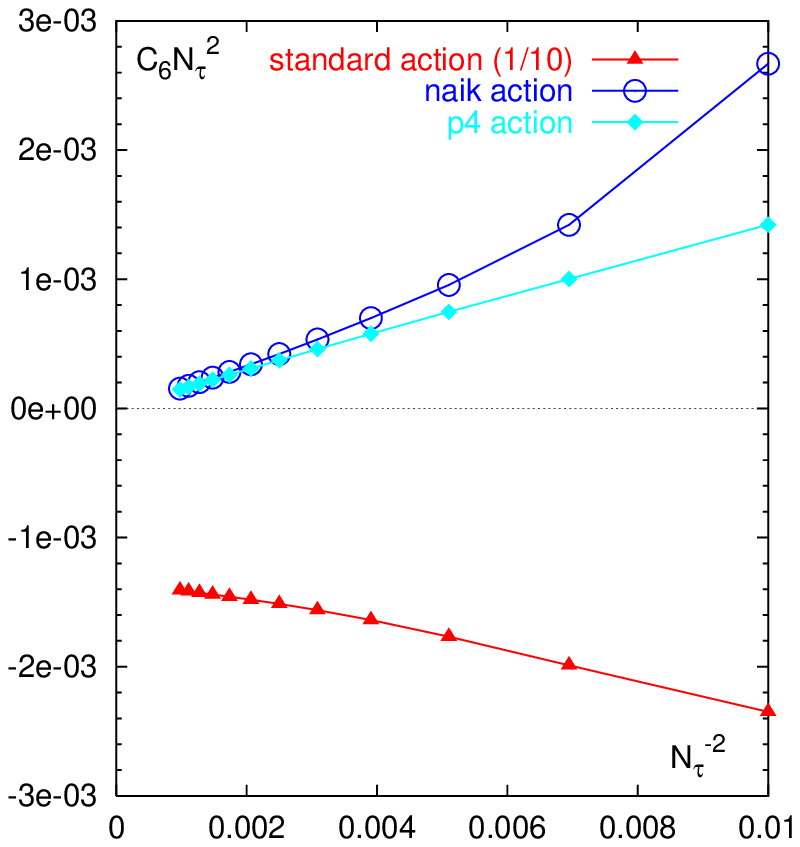, width=7.4cm}
(a)
\end{center}
\end{minipage}
\begin{minipage}[c][7.8cm][c]{7.4cm}
\begin{center}
\epsfig{file=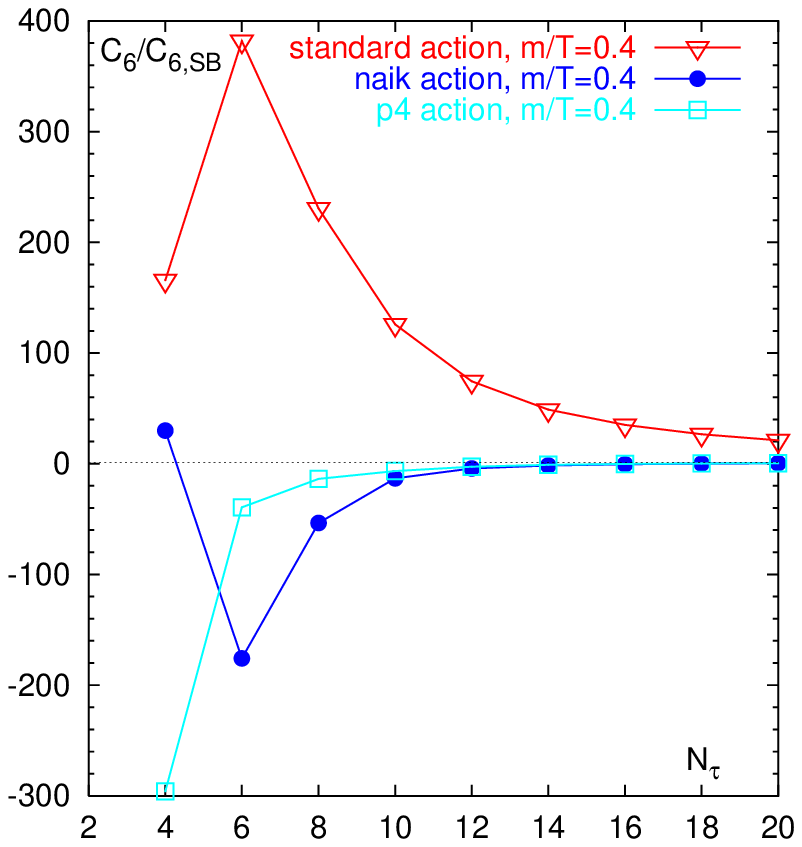, width=7.4cm}
(b)
\end{center}
\end{minipage}
\caption{(a) The coefficients ${\cal C}_6$ in the massless case, multiplied
  with $\Nt^2$ as a function of $\Nt^{-2}$ (results for standard
staggered fermions are divided by 10) and (b) 
the ratio ${\cal C}_6(N_\tau)/{\cal C}_6(\infty)$ at $m/T=0.4$ for the
    standard, Naik and p4 actions.}
\label{pressure_cutoff_mu_6}
\end{center}
\end{figure}
We see here again that the cutoff dependence of the pressure at $\mu\not=0$
is qualitatively the same as at $\mu=0$.

For massless quarks the coefficient ${\cal C}_6$ should
vanish with increasing $N_{\tau}$, as checked in
Fig.~\ref{pressure_cutoff_mu_6}a. It is expected that this term will
approach zero
like $N_{\tau}^{-2}$ in the large $N_{\tau}$ limit. In order to define the
numerical factor, we plot ${\cal C}_6N_\tau^2$ over $N_{\tau}^{-2}$. A fit
yields ${\cal C}_6\approx-0.015\,N_{\tau}^{-2}$ for the standard
action. This  is at least an order of magnitude larger than for the p4
improved action, for which the dominant cut-off dependence 
seems to be $O(N_\tau^{-4})$ as for the
Naik action.

In the case of massive quarks the expansion (\ref{eq:defcoeff}) no
longer terminates at ${\cal O}(\mu_q^4)$.
After expanding (\ref{pressure_con}) in terms of $\mu_q/T$ and
performing a numerical integration we find for the expansion
coefficients ${\cal C}_i (m/T)$ up to $i=6$ the values given in
Table~\ref{tab:table_pressure_massive}.
\begin{table}[tb]
\begin{center}
\begin{tabular}{|c||l|c|l|c|}\hline
& \multicolumn{2}{c|}{$m/T=0.4$} & \multicolumn{2}{c|}{$m/T=1.0$} \\ \hline
$i$ & ${\cal C}_i(m/T)$ & ${\cal C}_i(m/T)/{\cal C}_i(0)$ & ${\cal C}_i(m/T)$
& ${\cal C}_i(m/T)/{\cal C}_i(0)$ \\ \hline \hline
$0$ & $1.113632$              & 0.967 & $9.528163\cdot 10^{-1}$ & 0.827 \\
$2$ & $4.880455\cdot 10^{-1}$ & 0.976 & $4.313914\cdot 10^{-1}$ & 0.863 \\
$4$ & $2.531101\cdot 10^{-2}$ & 0.999 & $2.471397\cdot 10^{-2}$ & 0.976 \\
$6$ & $1.877659\cdot 10^{-6}$ &  ---  & $5.036816\cdot 10^{-5}$ & ---   \\ 
\hline
\end{tabular}
\caption{Continuum values for the coefficients ${\cal C}_i$ of the $\mu_q/T$
  expansion of the pressure of a massive gas of quarks for the mass values
  $m/T=0.4$ and $m/T=1.0$.}
\label{tab:table_pressure_massive}
\end{center}
\end{table}
The mass value $m/T=0.4$ is the value we use in our numerical calculations, 
corresponding to $\Nt=4$ and $am=0.1$.
We note that the coefficient ${\cal C}_6$ no longer vanishes.
As shown in Fig.~\ref{pressure_cutoff_mu_6}b, for $N_\tau$ finite there are
large deviations from the continuum value.
Even at $N_\tau=4$, however,
the absolute value of this coefficient is
still a factor of about $10^{-4}$ smaller than the leading term
${\cal C}_0$. 
The deviations thus do not show up in the calculation
of the complete expression for the pressure shown in
Figure~\ref{fig:pressure_cutoff_mu}a. These terms, however, become more
important in higher derivatives of the pressure such as the
quark number susceptibility $\chi_q$.
In summary, for a 
gas of free quarks we find that the $\mu_q/T$ expansion up to
$O((\mu_q/T)^4)$ is
sufficient for $\mu_q/T < 1$ and $m/T<1$. 
In the continuum the deviation from the full
expression over this range is smaller than 0.01\%. On the lattice, however,
cut-off
effects lead to deviations of approximately 10\% on coarse ($N_\tau=4$) 
lattices.

\section{Numerical Results}
\label{sec:results}

We applied the formalism outlined in Sec.~\ref{sec:form} to numerical
simulations of QCD with $N_{\rm f}=2$ quark flavors on a $16^3\times4$ lattice,
using both Symanzik improved gauge and p4-improved
staggered fermion actions. The simulation method is exactly as presented in
\cite{us} \footnote{The coefficient $c_3^F$ of the knight's move hopping term
was incorrectly reported to be 1/96 in \cite{us}; its correct value is 1/48.}.
The bare quark mass was $ma=0.1$ for which the
pseudocritical point for zero chemical potential is estimated to be
$\beta_c\simeq3.649(2)$. In order to cover a range of temperatures on
either side of the critical point we examined 16 values in the
range $\beta\in[3.52,4.0]$. The simulation employed a hybrid molecular dynamic
`R'-algorithm with discrete timestep $\delta\tau=0.025$, 
and measurements were performed on equilibrated
configurations separated by $\tau=5$. In general for each $\beta$ 500 to
800 configurations were analysed, with 1000 used in the critical region
$\beta\in[3.52,3.66]$. On each configuration
50 stochastic noise vectors were used to estimate the
required fermionic operators. For each noise vector, 7 matrix inversions 
are required to estimate the required operators 
(\ref{eq:dermu1}-\ref{eq:dermu4})
and (\ref{eq:45}-\ref{eq:48}).

Following the procedure used for the equation of state at $\mu=0$ \cite{KLP},
we translate to physical units using the following scaling ansatz \cite{Chris}:
\begin{equation}
{{a(\beta)}\over{a(\bar\beta)}}={\hat a}(\beta)
{{1+g_2\hat a^2(\beta)+g_4\hat a^4(\beta)}\over{1+g_2+g_4}}
\end{equation}
where $\hat a\equiv R(\beta)/R(\bar\beta)$, $R$ being the two-loop perturbative
scaling function appropriate for two light flavors. Using string tension data 
at $(T,\mu)=(0,0)$ the best values for the fit parameters corresponding to a 
reference
$\bar\beta=3.70$ are $g_2=0.669(208)$, $g_4=-0.0822(1088)$ \cite{KLP}. 
We find that 
our simulations span a temperature range $T/T_{0}\in[0.76,1.98]$, where 
$T_{0}$ is the critical temperature at $\mu_q=0$.

\begin{figure}[tb]
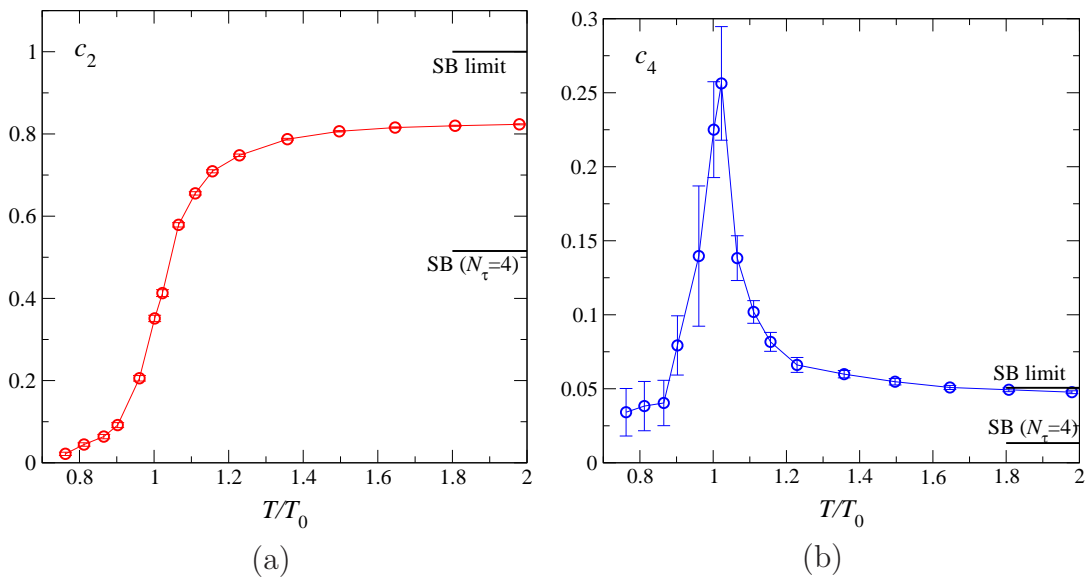

\begin{center}
\begin{minipage}[c][7.2cm][c]{7.2cm}
\begin{center}
\epsfig{file=c2vst_0412.eps, width=7.0cm}
\vspace{0.4cm}
(a)
\end{center}
\end{minipage}
\begin{minipage}[c][7.2cm][c]{7.2cm}
\begin{center}
\epsfig{file=c4vst_0412.eps, width=7.0cm}
\vspace{0.4cm}
(b)
\end{center}
\end{minipage}
\caption{Coefficients of (a) $(\mu_q/T)^2$ 
and (b) $(\mu_q/T)^4$ in the Taylor expansion of $\Delta(p/T^4)$
as functions of $T/T_{0}$.}
\label{fig:derivs}
\end{center}
\end{figure}
In Fig.~\ref{fig:derivs} we show the first two coefficients, (a) $c_2$ and
(b) $c_4$, of the Taylor expansion of $\Delta(p/T^4)$ introduced in
(\ref{eq:taylorcont})
as functions of $T/T_{0}$. Also shown are the 
corresponding SB limits: 
(a) $N_{\rm f}{\cal C}_2(N_\tau)$  and
(b) $N_{\rm f}{\cal C}_4(N_\tau)$, where the coefficients
${\cal C}_i$ are defined in (\ref{eq:defcoeff}), with values relevant for both 
the lattice used ($N_\tau=4$) 
and the continuum limit ($N_\tau=\infty$) plotted.
Both $c_2$ and $c_4$ vary sharply in the critical region, but
except in the immediate vicinity of the transition the quadratic term
dominates the quartic. This is consistent with the results of \cite{FKS}
where data at varying $\mu$ obtained by reweighting was found lie on an almost
universal curve when plotted as a fraction of the SB prediction.
The asymptotic value of $c_4$ appears to be approached from above.

A notable feature is that in the high-$T$ limit our data lies
closer to the continuum
SB prediction rather than their values ${\cal C}_i(N_\tau=4)$ 
corrected for lattice artifacts, $c_2$ assuming 80\% of the
continuum 
value for $T/T_{0}=2$ whereas $c_4$ is almost coincident with its
continuum value.
By contrast recent calculations with unimproved staggered
fermions \cite{FKS,GG} find that the high-$T$ limit of the data lie close to
the lattice-corrected SB value.
This situation can be modelled 
by making the coefficient $d$
of the $O(N_\tau^{-2})$ correction  appearing in (\ref{eq:cutcorr})
temperature dependent. In thermodynamic calculations performed with pure
unimproved SU(3) lattice gauge theory \cite{Boyd}, where extrapolations
to the continuum limit are currently practicable, it is found that
$d(T)\simeq 0.5d(T=\infty)$ for $T\sim 3T_{0}$, becoming even smaller closer to
$T_{0}$. The behaviour of $c_2$ and $c_4$ we have observed using
p4 fermions is broadly
consistent with this behaviour.

\begin{figure}
\begin{center}
\epsfig{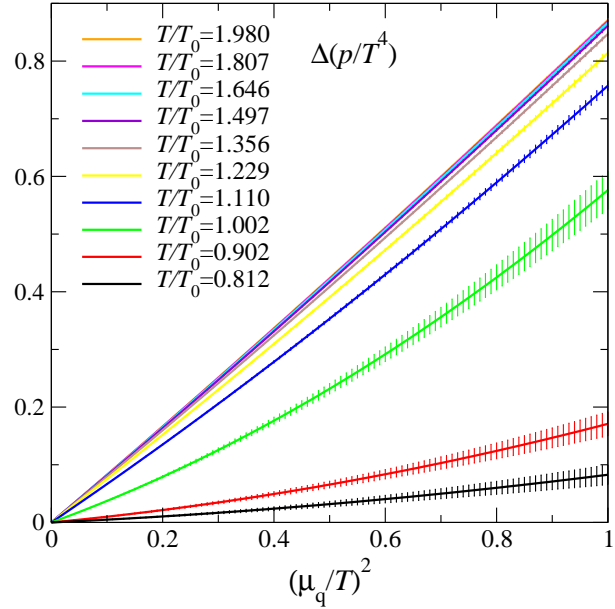}
\smallskip
\caption{
$\Delta(p/T^4)$ as a function of $(\mu_q/T)^2$ for
various temperatures, increasing upwards from
the lowest curve with $T/T_{0}=0.812$
to the highest with $T/T_{0}=1.980$.}
\label{fig:deltap}
\end{center}
\end{figure}

\begin{figure}
\begin{center}
\epsfig{file=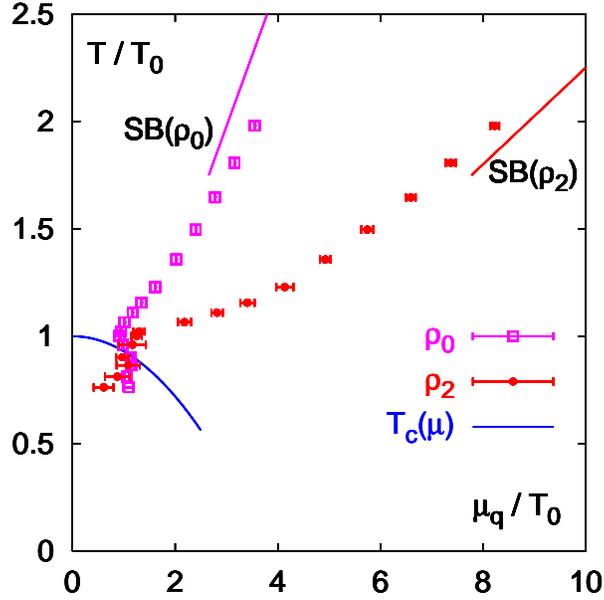, width=8.0cm}
\smallskip
\caption{
Estimates for the radius of convergence $\rho(\mu_q,T)$.}
\label{fig:rho}
\end{center}
\end{figure}

In Fig.~\ref{fig:deltap} we plot $\Delta(p/T^4)$ defined in 
(\ref{eq:taylorcont}) as a function of $(\mu_q/T)^2$
for various temperatures.
In most cases $c_4\ll c_2$
and the
relation thus almost linear. The strongest departures from linearity are for
$T\simeq T_{0}$, but even here the quadratic term is dominant for
$(\mu_q/T)^2\lsim0.4$, corresponding to $\mu_q\lsim100$MeV. 
Given enough terms of the Taylor expansion in $\mu_q/T$, one could determine 
its radius of convergence $\rho$ via\footnote{The argument of \cite{dFP} that 
$\rho\leq{\pi\over3}$ due to the presence of a phase transition as imaginary
chemical potential is increased beyond this value \cite{roberge} does not hold
for calculations with $\mu_q$ real; in this case the pressure $p_0$
corresponding to the unit element of the Z(3) sector is always the maximum and 
hence dominates the partition function in the thermodynamic limit -- hence the
issue concerns the analytic properties 
{\em within} this physical unit sector.} 
\begin{equation}
\rho=\lim_{n\to\infty}\rho_n\equiv\lim_{n\to\infty}\sqrt{\left\vert
{c_n\over c_{n+2}}\right\vert}.
\end{equation}
Data from the pressure at $\mu_q=0$ \cite{KLP} and the current study enable us 
to plot the first two estimates $\rho_0$ and $\rho_2$ on the $(\mu_q,T)$
plane along with the estimated pseudocritical line
$T_c(\mu_q)$ found in \cite{us} in Fig.~\ref{fig:rho}.
Also shown are the corresponding values from the SB limit 
(\ref{Stefan-Boltzmann}). For $T>T_c$ one finds that $\rho_n$ 
increases markedly as $n$ increases from 0 to
2; if the SB limit is a good predictor for the QGP phase we might expect
$c_6$ to be very small, and the next estimate $\rho_4$ correspondingly very
large in this regime. Close to the transition line, however, the thermodynamic
singularities appear to restrict $\rho\sim O(1)$; this in turn gives 
an approximate lower bound for the position of the critical endpoint.
From the figure we deduce
$\mu_q^{crit}\gsim(1-1.2)T_{0}$, not inconsistent with the result of 
\cite{FK}.
The new results at $O(\mu^4)$ are important
because they justify in retrospect our neglect of fourth order reweighting
factors in our earlier calculation of the critical line $T_c(\mu_q)$ \cite{us}.
Indeed, simulations with imaginary $\mu$ suggest that neglect of these terms in
the analytic continuation to physical $\mu_q$ is
justified for $\mu_q\lsim170$MeV \cite{dFP}.

\begin{figure}[tb]
\begin{center}
\begin{minipage}[c][7.4cm][c]{7.4cm}
\begin{center}
\epsfig{file=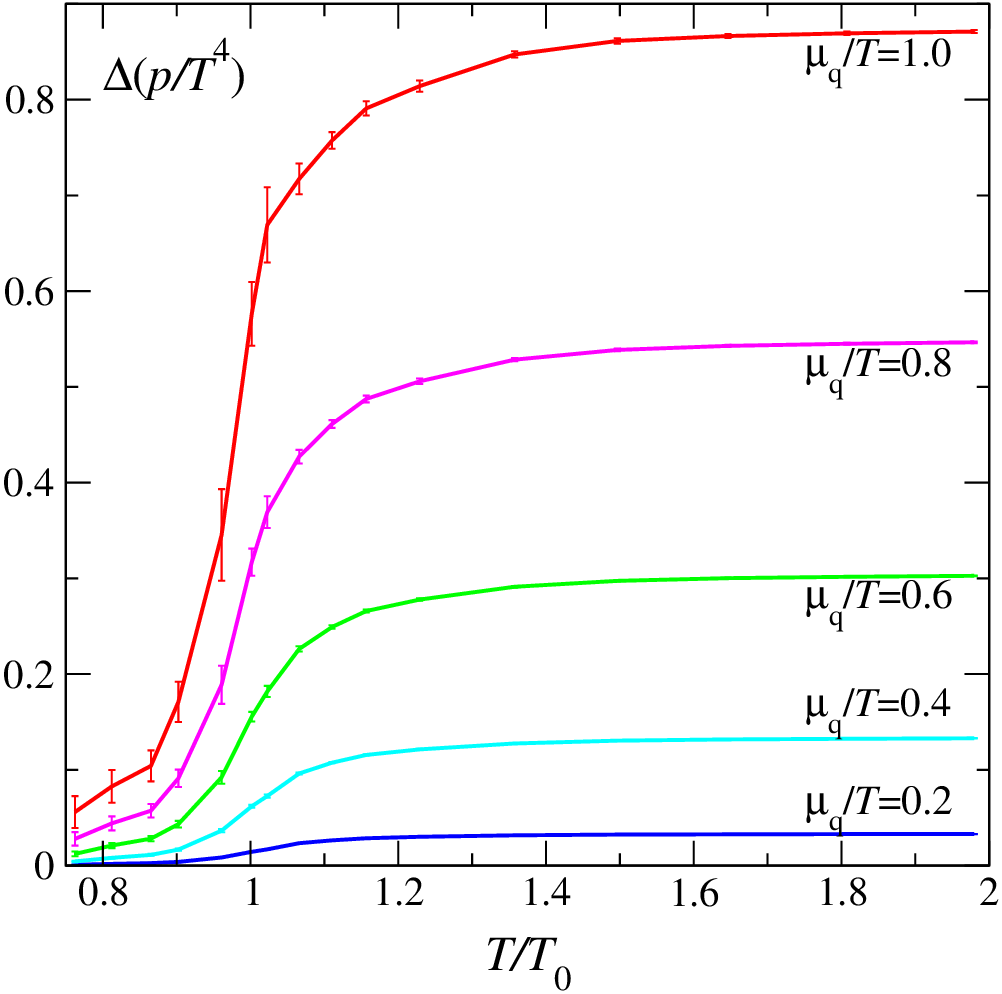, width=7.4cm}
\vspace{0.4cm}
(a)
\end{center}
\end{minipage}
\begin{minipage}[c][7.4cm][c]{7.4cm}
\begin{center}
\epsfig{file=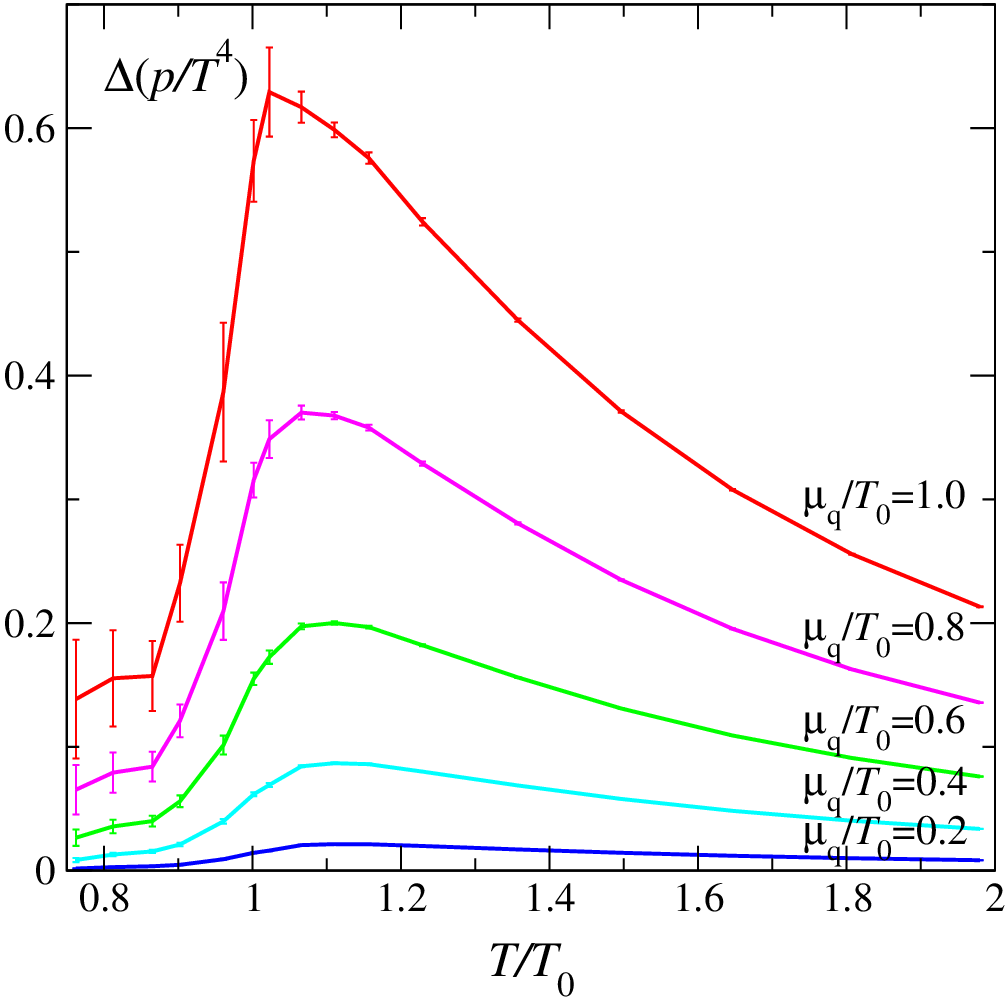, width=7.4cm}
\vspace{0.4cm}
(b)
\end{center}
\end{minipage}
\caption{The equation of state correction $\Delta(p/T^4)$ vs. $T/T_{0}$
for (a) various $\mu_q/T$, and (b) various $\mu_q/T_{0}$.}
\label{fig:eos}
\end{center}
\bigskip\bigskip
\begin{center}
\begin{minipage}[c][7.4cm][c]{7.4cm}
\begin{center}
\epsfig{file=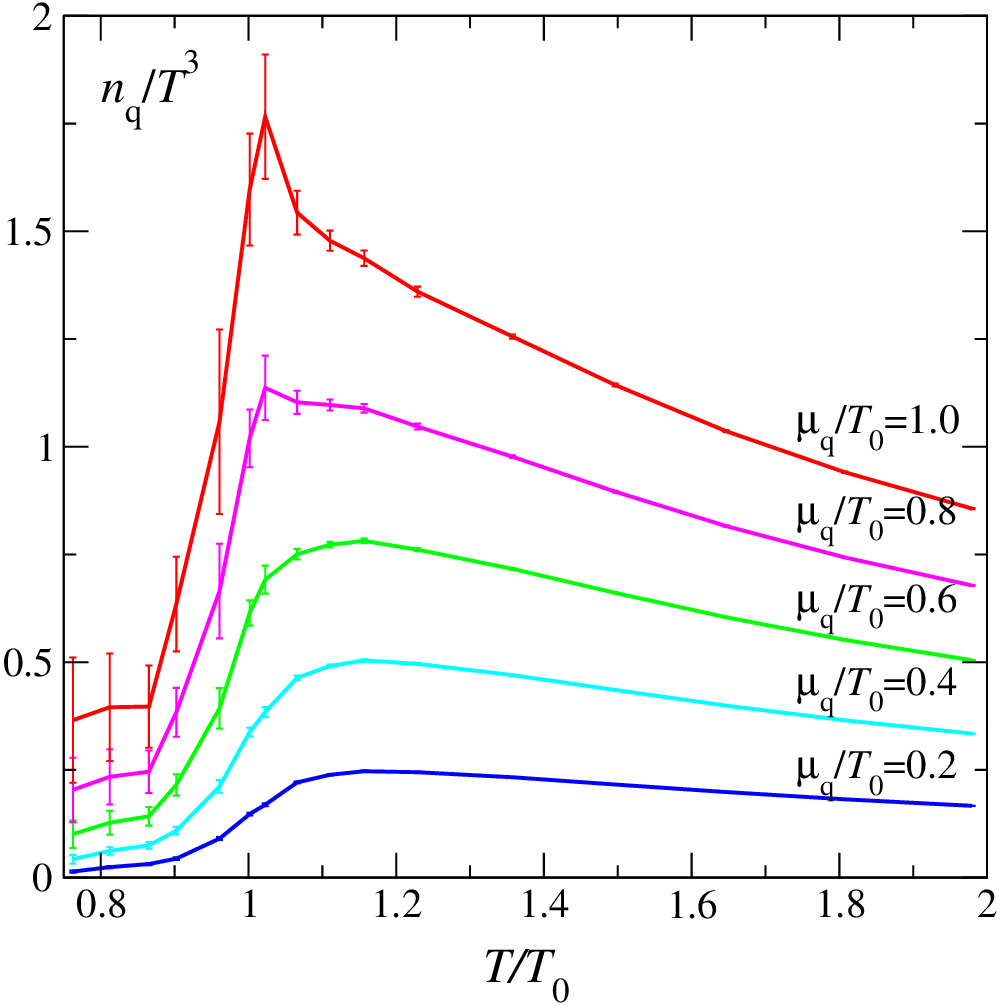, width=7.4cm}
\vspace{0.4cm}
(a)
\end{center}
\end{minipage}
\begin{minipage}[c][7.4cm][c]{7.4cm}
\begin{center}
\epsfig{file=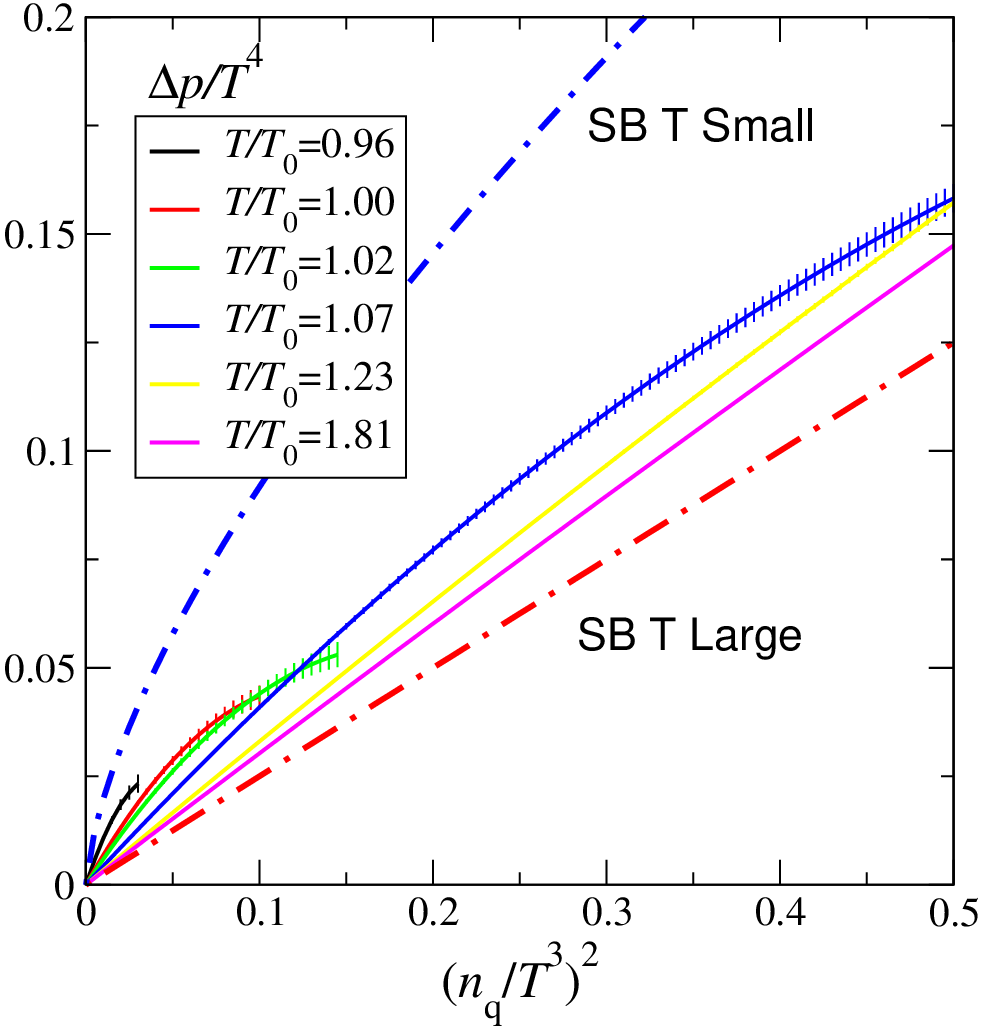, width=7.4cm}
\vspace{0.4cm}
(b)
\end{center}
\end{minipage}
\caption{(a) $n_q/T^3$ as a function of $T/T_{0}$ for
various $\mu_q/T_{0}$,
and (b) 
the ''true'' equation of state $\Delta(p/T^4)$ vs. $(n_q/T^3)^2$
for various temperatures. The continuum SB forms
${1\over4}(\pi^2/N_{\rm f})^{1/3}(n_q/T^3)^{4/3}$ (low $T$) and
$(2N_{\rm f})^{-1}(n_q/T^3)^2$ (high $T$) are also shown
as functions of $T/T_{0}$.}
\label{fig:density}
\end{center}
\end{figure}

Fig.~\ref{fig:eos} plots the dimensionless
correction $\Delta(p/T^4)$ to the equation of state as a function of both 
$\mu_q/T$ and $\mu_q$. In the latter case 
the correction rises steeply across the transition and
peaks for $T\simeq1.1T_{0}$, before rapidly approaching a form
$\Delta(p/T^4)=\alpha T^{-2}$ characteristic of the SB
limit, with the coefficient $\alpha$ having 82\% of the continuum SB value.
Comparison with the equation of state results at $\mu_q=0$ from
Ref.~\cite{KLP} suggest that the correction will give a significant correction
to the pressure for $0.9\lsim T/T_{0}\lsim1.3$, $\mu_q/T_{0}\gsim0.5$,
but will decrease in importance as $T$ rises further.
The curves of Fig.~\ref{fig:eos}b are in good qualitative agreement with
those of Refs.~\cite{FKS,GG}, although we consider 
any quantitative agreement to be somewhat accidental as the numerical data 
obtained in \cite{FKS,GG}  with unimproved actions have large
discretisation errors which have been corrected for by renormalising
the raw data with the known discretisation errors in the
infinite temperature limit. Experience gained in calculations
of thermodynamic quantities in the pure SU(3) gauge theory suggests
that in the temperature range of a few times $T_{0}$ this procedure
overestimates the importance of cut-off effects by a factor two or so
\cite{Boyd}.

Fig.~\ref{fig:density}a shows the quark number density $n_q$ evaluated using 
Eqn.~(\ref{eq:density}). As $\mu_q$ increases, $n_q$
rises steeply
as the QGP phase is entered; for reference, 
if the quark number density in nuclear
matter is denoted $\bar n_q$, 
then the ratio $\bar n_q/T_{0}^3\approx0.75$. Our
results are numerically very similar to those obtained using exact reweighting
in \cite{FKS}, where a mass $ma\approx0.1$ for the light quark
flavors was used. Note that a significant quark mass dependence 
for $n_q$ was observed in \cite{us},  and indeed is present 
even in the SB limit as described in Section~\ref{cut-off_dependence};
however analysis of the 
SB limit suggests that the difference between the chiral
limit and $m/T=0.4$
is about 4\%. In Fig.~\ref{fig:density}b we show the result of eliminating
$\mu_q$ in favour of $n_q$ via
\begin{equation}
\Delta\left({p\over T^4}\right)={1\over{4c_2}}\left({n_q\over T^3}\right)^2
-{3c_4\over16c_2^4}\left({n_q\over T^3}\right)^4
+O\biggl(\left({n_q\over T^3}\right)^6\biggr).
\label{eq:eostrue}
\end{equation}

\clearpage

The relation (\ref{eq:eostrue}) approximates the
``true'' equation of state in
terms of physically measurable quantities; we have plotted the resulting
$\Delta(p/T^4)$ against $(n_q/T^3)^2$ up to the point where the ratio
of the magnitude of the second term
of (\ref{eq:eostrue}) to that of the first is 40\%: the point
$n_q/T^3=\sqrt{2c_2^3/3c_4}$
where the ratio is 50\% marks a mechanical instability
$\partial p/\partial n_q=0$, which is an artifact due to the truncation
of the
series. Stability of the equilibrium state under local fluctuations
$\delta n_q$
requires $\partial p/\partial n_q>0$, an example of
Le Ch\^atelier's principle.
As $T/T_{0}$ increases through unity, the equation of state
changes from a form resembling the low-$T$ SB limit $p\propto n_q^{4/3}$
to the stiffer $p\propto n_q^2$ characteristic of the high-$T$ SB limit.
Interestingly enough, to the order we have calculated
the instability artifact sets in at
$\mu_q/T\simeq1.4$ for $T$ large, but at
$\mu_q/T\simeq0.4$ for $T\approx T_{0}$, thus providing an independent,
and
more stringent, limit to the physical validity of our approach, and
reflecting
the importance of contributions from higher orders in the Taylor
expansion
close to $T_c(\mu_q)$. 

\begin{figure}[tb]
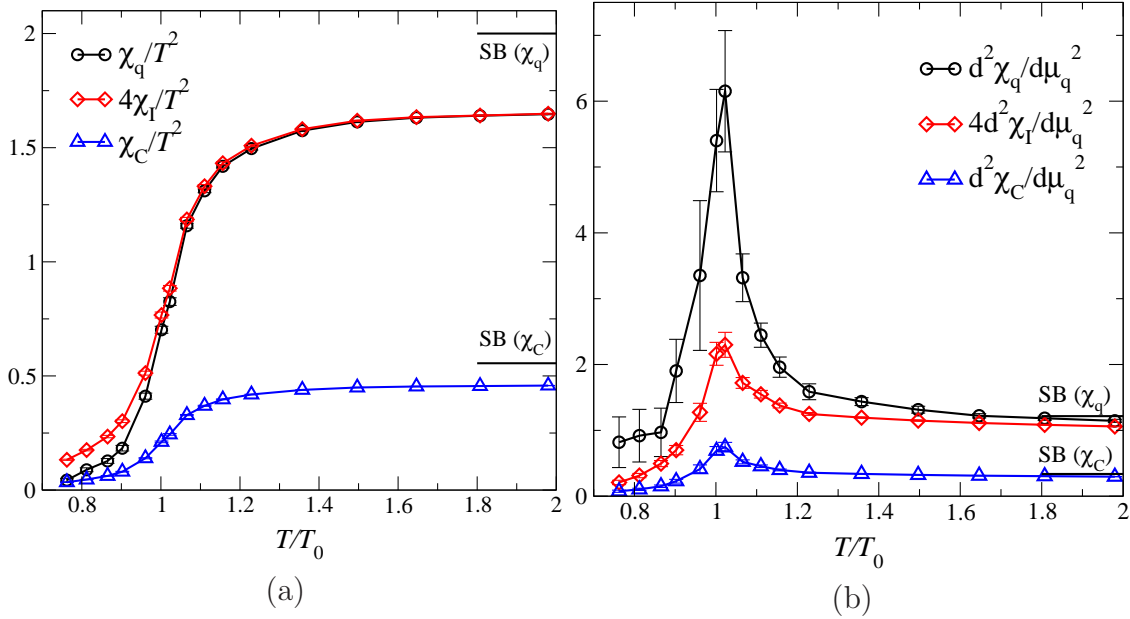

\begin{center}
\begin{minipage}[c][7.4cm][c]{7.4cm}
\begin{center}
\epsfig{file=qns_0423.eps, width=7.4cm}
\vspace{0.4cm}
(a)
\end{center}
\end{minipage}
\begin{minipage}[c][7.4cm][c]{7.4cm}
\begin{center}
\epsfig{file=dqnsdmu2_0423.eps, width=7.4cm}
\vspace{0.4cm}
(b)
\end{center}
\end{minipage}
\caption{Susceptibilities (a) 
$\chi_i/T^2\vert_{\mu_q=0}$,  and (b)
$\partial^2\chi_i/\partial\mu_q^2\vert_{\mu_q=0}$
as functions of $T/T_{0}$.}
\label{fig:chistc}
\end{center}
\end{figure}
Next,  in Fig.~\ref{fig:chistc} 
we plot the expansion coefficients corresponding to
the various susceptibilities defined 
in (\ref{eq:chisq}-\ref{eq:chic}). For $T\lsim T_{0}$ there is a 
significant difference between $\chi_q(\mu_q=0)$ and 
$4\chi_{\rm I}(\mu_q=0)$, 
implying anti-correlated fluctuations of $n_{\rm u}$ and $n_{\rm d}$ 
which rapidly decrease in magnitude above $T_0$ and
vanish as $T$ approaches the infinite temperature 
SB limit\footnote{There has recently been a discussion
whether the difference $(4\chi_I - \chi_q)/T^2$ is exactly zero in the 
high temperature phase, as suggested by some lattice calculations
\cite{gavai}, or just small but non-zero, as suggested by HTL-resummed
perturbation theory \cite{rebhan}. We find that the difference stays
non-zero but decreases by one order of magnitude between $T\simeq T_0$ 
and $T\simeq 1.5 T_0$.  
At $T\simeq 1.36 T_0$ we find a value of 0.0066(28) for this difference
calculated in 2-flavor QCD which clearly disagrees with the quenched
result $(2\pm 4)\cdot 10^{-6}$ presented in \cite{gavai} as well as
the recent 2-flavor results of this group \cite{GG}. Our results
are, however, in agreement with the findings of Ref.~\cite{milc}.
At $T\simeq 2T_0$ the numerical value of this difference drops
below our current error level of about $2\cdot 10^{-3}$.
In the high temperature limit this error is thus not yet small enough 
to discuss numerical effects at the level of $10^{-4}$ as suggested
in the discussion presented in \cite{blaizot}.}.
In the same limit the charge susceptibility $\chi_{\rm C}$ approaches the
value ${5\over18}\chi_q$. The critical 
singularity in $4\chi_{\rm I}$ and $\chi_{\rm C}$ is weaker
than that of $\chi_q$, which can be traced back to the differing
coefficients of 
$\langle(\partial^2\ln\mbox{det}M/\partial\mu^2)^2\rangle$, the dominant term
in the vicinity of $T_c$, 
in the definitions  (\ref{eq:d4lnz}) and (\ref{eq:chit}).
The dimensionless quantity $T\chi_{\rm C}/s$,
where $s=(\epsilon+p-\mu_qn_q)/T$ is the entropy density,
can be related to event-by-event fluctuations 
in charged particle multiplicities
in RHIC collisions, and 
has been proposed as a signal for
QGP formation \cite{AHMJK}. Event-by-event fluctuations in baryon number have 
also recently been discussed in \cite{HS}.

\begin{figure}
\begin{center}
\epsfig{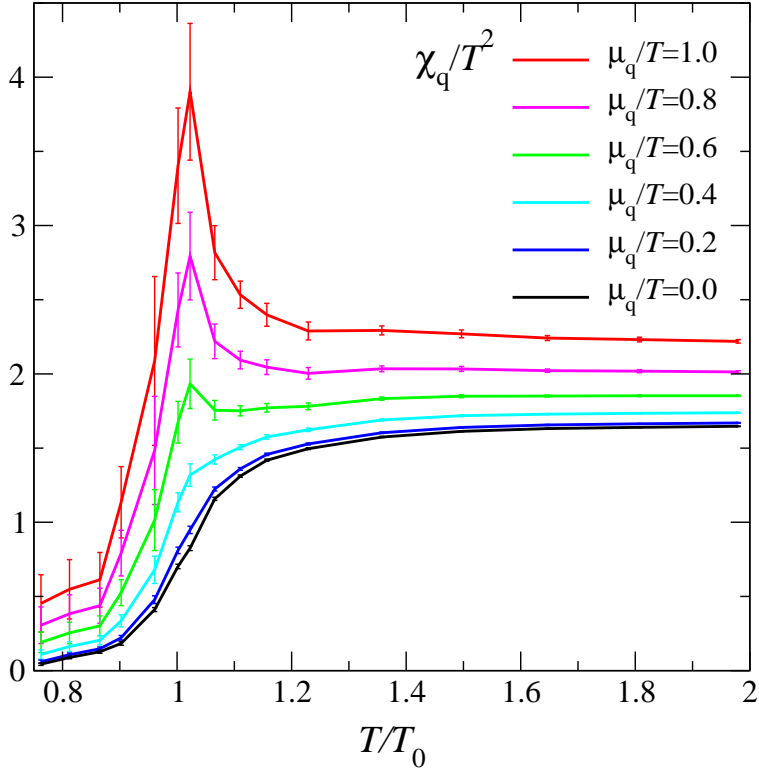}
\smallskip
\caption{$\chi_q/T^2$ as a function of $T/T_{0}$ for
various $\mu_q/T$.
}
\label{fig:chis}
\end{center}
\end{figure}

In Fig.~\ref{fig:chis}
the relation (\ref{eq:susc}) and data of Fig.~\ref{fig:derivs}
have been used to plot the dimensionless
quark number
susceptibility $\chi_q/T^2$ as a function of $T/T_{0}$ for various $\mu_q/T$.
The peak which develops in $\chi_q$ as $\mu_q$ increases is a sign that
fluctuations in the baryon density are growing as the critical endpoint in the
$(\mu,T)$ plane is approached. Physically, this shows that at the critical
point, as well as strong fluctuations in the $(\bar\psi\psi)$ bilinear expected
at a chiral phase transition there are also fluctuations in
$(\bar\psi\gamma_0\psi)$ since Lorentz symmetry is explicitly broken by the
background baryon charge density. 
For quantities such as $n_q$ and $\chi_q$ defined as higher derivatives of the 
free energy with respect to $\mu_q$, 
the relative importance of the higher order
terms in the Taylor series expansion is increased; for example, at $T\simeq
T_{0}$
and $\mu_q/T=1$ the quadratic contribution to $\chi_q(\mu_q)$
is about 3 times that of the leading order term.
For this reason we do not expect the data of Fig.~\ref{fig:chis}
to be quantitatively accurate in the critical region.
Note, however, that at each temperature 
the expansions for $p$, $n_q$ and $\chi_q$ all have the 
{\em same\/} radius of convergence.

\begin{figure}[tb]
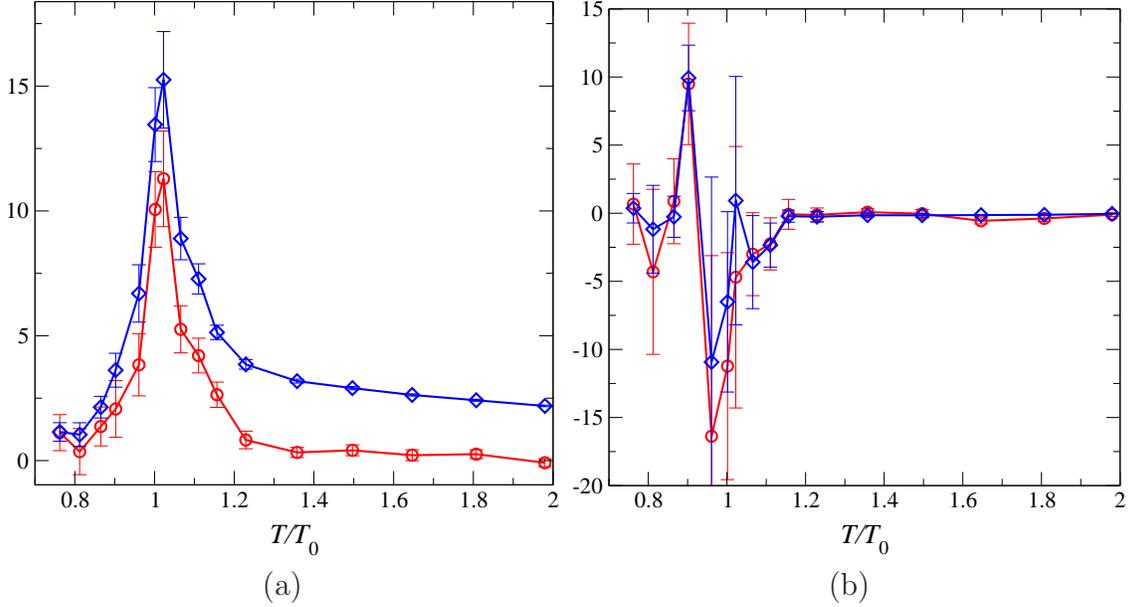

\bigskip\bigskip\bigskip
\begin{center}
\begin{minipage}[c][7.4cm][c]{7.4cm}
\begin{center}
\epsfig{file=d3pdmu2dbm_0429.eps, width=7.4cm}
\vspace{0.4cm}
(a)
\end{center}
\end{minipage}
\begin{minipage}[c][7.4cm][c]{7.4cm}
\begin{center}
\epsfig{file=d5pdmu4dbm_0429.eps, width=7.4cm}
\vspace{0.4cm}
(b)
\end{center}
\end{minipage}
\caption{Derivatives (a) $(2VT^3)^{-1}\partial^3\ln{\cal
Z}/\partial\beta\partial(\mu_q/T)^2$ (circles) and 
$-(2VT^3)^{-1}\partial^3\ln{\cal 
Z}/\partial m\partial(\mu_q/T)^2$ (diamonds); (b) 
$(24VT^3)^{-1}\partial^5\ln{\cal
Z}/\partial\beta\partial(\mu_q/T)^4$ (circles) and 
$-(24 VT^3)^{-1}\partial^5\ln{\cal 
Z}/\partial m\partial(\mu_q/T)^4$ (diamonds).}
\label{fig:e-3p}
\end{center}
\end{figure}

Finally we turn to a discussion of the derivatives necessary for calculating
the response of the energy density $\epsilon$ to increasing $\mu_q$.
The Taylor expansion of the energy density involves derivatives of the
expansion coefficients $c_p(T)$ used to calculate the pressure,
\begin{equation}
\Delta\left({{\epsilon-3p}\over T^4}\right) = \sum_{p=1}^\infty c_p'(T)
\biggl( {\mu_q \over T} \biggr)^{p} \quad ,
\label{epsTaylor}
\end{equation}
with $c_p'(T) = T({\rm d}c_p(T)/{\rm d}T)\vert_{\mu_q=0}$. It is apparent from 
the temperature dependence of the expansion coefficients $c_2(T)$ and
$c_4(T)$ shown in Fig.~\ref{fig:derivs} that the coefficients $c_p'(T)$
can become large in the vicinity of $T_0$. On the other hand that
figure also shows that $c_p'(T)$ will be small, {\it i.e.} close to
zero, at high temperature as expected in the ideal gas limit.
A comparison with (\ref{eq:deltaeps}) shows that the numerical evaluation
of $c_p'(T)$ requires the knowledge of lattice beta-functions and a
calculation of mixed derivatives of
$\ln {\cal Z}$ with respect to $\mu$ as well as $\beta$ and $m$.
In Fig.~\ref{fig:e-3p} we plot these derivative terms;
the signals in this case are much noisier than for
$\partial^n\ln{\cal Z}/\partial\mu^n$, although we have been able to
check that the numerical values for $\partial^3\ln{\cal
Z}/\partial\beta\partial\mu^2$
are consistent with the slope of the curve in Fig.~\ref{fig:derivs}a. It
is
clear firstly that with the exception of $\partial^3\ln{\cal Z}/\partial
m\partial\mu^2$ the signal only differs significantly from zero
in the immediate neighbourhood of the transition, and secondly
that derivatives with
respect to $m$ are strongly anti-correlated with those with respect to
$\beta$.
The latter suggests it might be possible to learn something from
(\ref{eq:deltaeps}) about the shape of the
$\Delta((\epsilon-3p)/T^4)$ curve
as a function of $T/T_0$ away from the chiral limit even in the absence
of quantitative  information about $a\partial m/\partial a$.

Consider however ignoring mass derivatives and focusing on 
those performed  with respect to coupling. In this case all derivatives
are consistent with zero for $T\gsim 1.2\;T_0$; {\it i.e.} the 
difference $\Delta\left({{(\epsilon-3p)}/T^4}\right)$ is to a good
approximation independent of $\mu_q$ for these high temperatures. This
observation is consistent with the results obtained by using exact 
reweighting in \cite{FKS}.  Consider now temperatures close to $T_0$.
The beta
function at the critical $\beta_c$ has the value $a^{-1}da/d\beta=-2.08(43)$
\cite{KLP}; substituting the derivatives from Fig.~\ref{fig:e-3p}
in (\ref{eq:deltaeps}) we find at $T_0$
\begin{equation}
\Delta\left({{\epsilon-3p}\over
T^4}\right)\approx(4.8\pm1.2)\times\left({\mu_q\over T}\right)^2-(5\pm4)\times
\left({\mu_q\over T}\right)^4+\cdots
\label{eq:e-3p}
\end{equation}
Taking the central values of the coefficients in this expansion one may
conclude that the ratio $c_2'/c_4'$ is comparable with $c_2/c_4$.
At present the large error on the coefficient of the $(\mu_q/T)^4$ term,
however, does not allow a firm conclusion on the convergence
radius of the expansion of $\epsilon-3p$. We also note that the
coefficient $c_4'$ will change sign for $T\sim T_0$. This suggests 
that large cancellations can occur for
$\mu_q/T \sim {\cal O}(1)$ and indicates that higher order terms
are needed to determine this difference reliably.
In any event, it would appear that extending our current analysis
to determine energy and entropy densities ($\epsilon$, $s$) in the 
critical region will be far more demanding.

\section{Summary}
\label{sec:andfinally}

We have presented the first Monte Carlo calculation of the QCD
equation of state at non-zero quark chemical potential within the analytic
framework; no reweighting has been performed. As in our previous work, we have
exploited the relative simplicity of the method
to explore larger physical volumes than those used in 
comparable studies \cite{FKS,GG}. In addition, the compatibility of our method
with the use of an improved lattice fermion action has meant that our results
suffer from relatively mild discretisation artifacts, our data for the 
pressure correction $\Delta p(\mu_q)$
achieving 80\% of the Stefan-Boltzmann value by  $T\simeq2T_{0}$.

Our results for $\Delta p$ and its $\mu$-derivatives $n_q$ and the various
susceptibilities $\chi_i$ are in good qualitative agreement with those of 
\cite{FKS,GG}. Since higher derivatives suffer from larger discretisation
artifacts, and are inherently noisier when estimated by Monte Carlo simulation,
the results for, say, $\chi_q$ are less quantitatively reliable than those
for $\Delta p$; nonetheless the singularity developing in $\chi_q$ as $\mu_q$
is increased, seen in Fig.~\ref{fig:chis}, is evidence for 
the presence of a critical endpoint in the
$(\mu_q,T)$ plane, and for the importance of quark number fluctuations in its
vicinity.

The calculation of fourth order derivatives has enabled us for the first
time to make an estimate for the limitations of our method, both analytically
through the radius of convergence of the Taylor expansion in $\mu_q/T$, and
physically via the requirement of consistency with Le Ch\^atelier's principle.
For both criteria the most stringent bounds are unsurprisingly in the vicinity 
of the critical line, where convergence of the series limits us to
$\mu_q/T\lsim1$, and mechanical stability of the equilibrium state to
$\mu_q/T\lsim0.5$. Of course, the picture should change with
the inclusion of still higher derivatives since on physical grounds we expect 
stability of the equilibrium state everywhere within the domain of convergence.
We are currently investigating the feasibility of including the relevant
$O(\mu^6)$ terms in our calculation.

Other quantities of phenomenological importance such as the energy $\epsilon$
and entropy $s$ densities, which require mixed derivatives with respect to the
other bare parameters $\beta$ and $m$,
appear more difficult to calculate with quantitative 
accuracy with this approach. It remains an open question whether the radius 
of convergence for these quantities is the same as that for quantities
defined by series in $\partial^n\Omega/\partial\mu^n$.

Finally, it is necessary to stress the importance of refining the current
calculation, firstly 
by simulating systems with $N_\tau\geq6$ so that a
reliable extrapolation to the continuum can be performed,
and secondly by repeating it with a realistic spectrum of 2+1 fermion
flavors. 

\section*{Acknowledgements}
Numerical work was performed using a 128-processor APEMille in Swansea,
supported by PPARC grant PPA/G/S/1999/00026; this grant also supported the work
of SE. SJH is supported by a PPARC Senior Research Fellowship. 
Work of the Bielefeld group has been supported partly through the DFG
under grant FOR 339/2-1 and a grant of the BMBF under contract no.
06BI106.
We thank 
Zoltan Fodor, S\'andor Katz and Krishna Rajagopal for helpful discussions.


\begin{appendix}
\section{Derivatives needed to calculate energy density}
\label{appendix:eps}
Here we present the non-vanishing terms in the expressions for 
$\partial^{n+1}\ln{\cal Z}/\partial m\partial\mu^n$:
\begin{eqnarray}
\label{eq:d3zdmu2dm}
&&\hspace{-2cm}\frac{\partial^3 \ln {\cal Z}}{\partial \mu^2 \partial m}
\hspace{1.2cm}=
\left\langle \frac{N_{\rm f}}{4}
\frac{\partial^2 {\rm tr} M^{-1}}{\partial \mu^2} \right\rangle
+ 2 \left\langle \left( \frac{N_{\rm f}}{4} \right)^2
\frac{\partial (\ln \det M)}{\partial \mu}
\frac{\partial {\rm tr} M^{-1}}{\partial \mu}  \right\rangle \\
&&+ \left\langle \left( \frac{N_{\rm f}}{4} \right)^2
\frac{\partial^2 (\ln \det M)}{\partial \mu^2} {\rm tr} M^{-1}
\right\rangle
+ \left\langle \left( \frac{N_{\rm f}}{4} \right)^3
\left( \frac{\partial (\ln \det M)}{\partial \mu} \right)^2
{\rm tr} M^{-1} \right\rangle
\nonumber \\
&&- \left[ \left\langle \frac{N_{\rm f}}{4}
\frac{\partial^2 (\ln \det M)}{\partial \mu^2} \right\rangle
+ \left\langle \left( \frac{N_{\rm f}}{4}
\frac{\partial (\ln \det M)}{\partial \mu} \right)^2 \right\rangle
\right]
\left\langle \frac{N_{\rm f}}{4} {\rm tr} M^{-1} \right\rangle
\nonumber\\
\label{eq:d5zdmu4dm}
&&\hspace{-2cm}\frac{\partial^5 \ln {\cal Z}}{\partial \mu^4 \partial m}
\hspace{1.2cm}=
\left\langle \frac{N_{\rm f}}{4}
\frac{\partial^4 ({\rm tr} M^{-1})}{\partial \mu^4} \right\rangle
+ 4 \left\langle \left( \frac{N_{\rm f}}{4} \right)^2
\frac{\partial^3 ({\rm tr} M^{-1})}{\partial \mu^3}
\frac{\partial (\ln \det M)}{\partial \mu} \right\rangle \\
&& \hspace{-1.2cm}
+ 4 \left\langle \left( \frac{N_{\rm f}}{4} \right)^2
\frac{\partial^3 (\ln \det M)}{\partial \mu^3}
\frac{\partial ({\rm tr} M^{-1})}{\partial \mu} \right\rangle
+ 6 \left\langle \left( \frac{N_{\rm f}}{4} \right)^2
\frac{\partial^2 (\ln \det M)}{\partial \mu^2}
\frac{\partial^2 ({\rm tr} M^{-1})}{\partial \mu^2}
\right\rangle \nonumber \\
&& \hspace{-1.2cm}
+ 6 \left\langle \left( \frac{N_{\rm f}}{4} \right)^3
\frac{\partial^2 ({\rm tr} M^{-1})}{\partial \mu^2}
\left( \frac{\partial (\ln \det M)}{\partial \mu} \right)^2
\right\rangle\nonumber\\
&& \hspace{-1.2cm}
+ 12 \left\langle \left( \frac{N_{\rm f}}{4} \right)^3
\frac{\partial^2 (\ln \det M)}{\partial \mu^2}
\frac{\partial (\ln \det M)}{\partial \mu}
\frac{\partial ({\rm tr} M^{-1})}{\partial \mu}
\right\rangle \nonumber \\
&& \hspace{-1.2cm}
+ 4 \left\langle \left( \frac{N_{\rm f}}{4} \right)^4
\left( \frac{\partial (\ln \det M)}{\partial \mu} \right)^3
\frac{\partial ({\rm tr} M^{-1})}{\partial \mu}
\right\rangle 
+ \left\langle \left( \frac{N_{\rm f}}{4} \right)^2
\frac{\partial^4 (\ln \det M)}{\partial \mu^4}
{\rm tr} M^{-1} \right\rangle\nonumber\\
&& \hspace{-1.2cm}
+ 4 \left\langle \left( \frac{N_{\rm f}}{4} \right)^3
\frac{\partial^3 (\ln \det M)}{\partial \mu^3}
\frac{\partial (\ln \det M)}{\partial \mu} {\rm tr} M^{-1}
\right\rangle \nonumber \\
&& \hspace{-1.2cm}
+ 3 \left\langle \left( \frac{N_{\rm f}}{4} \right)^3
\left( \frac{\partial^2 (\ln \det M)}{\partial \mu^2}
\right)^2 {\rm tr} M^{-1} \right\rangle\nonumber\\
&& \hspace{-1.2cm}
+ 6 \left\langle \left( \frac{N_{\rm f}}{4} \right)^4
\frac{\partial^2 (\ln \det M)}{\partial \mu^2}
\left( \frac{\partial (\ln \det M)}{\partial \mu} \right)^2
{\rm tr} M^{-1} \right\rangle \nonumber \\
&& \hspace{-1.2cm}
+ \left\langle \left( \frac{N_{\rm f}}{4} \right)^5
\left( \frac{\partial (\ln \det M)}{\partial \mu} \right)^4
{\rm tr} M^{-1} \right\rangle \nonumber \\
&& \hspace{-1.2cm}
- \Biggl[ \left\langle \frac{N_{\rm f}}{4}
\frac{\partial^4 (\ln \det M)}{\partial \mu^4} \right\rangle
+ 4 \left\langle \left( \frac{N_{\rm f}}{4} \right)^2
\frac{\partial^3 (\ln \det M)}{\partial \mu^3}
\frac{\partial (\ln \det M)}{\partial \mu} \right\rangle
\nonumber \\
&& + 3 \left\langle \left( \frac{N_{\rm f}}{4} \right)^2
\left( \frac{\partial^2 (\ln \det M)}{\partial \mu^2}
\right)^2 \right\rangle
+ \left\langle \left( \frac{N_{\rm f}}{4}
\frac{\partial (\ln \det M)}{\partial \mu} \right)^4 \right\rangle
\nonumber\\
&&
+ 6 \left\langle \left( \frac{N_{\rm f}}{4} \right)^3
\frac{\partial^2 (\ln \det M)}{\partial \mu^2}
\left( \frac{\partial (\ln \det M)}{\partial \mu} \right)^2
\right\rangle\Biggr]
\left\langle \frac{N_{\rm f}}{4} {\rm tr} M^{-1} \right\rangle
 \nonumber \\
&& \hspace{-1.2cm}
- 6 \left[ \left\langle \frac{N_{\rm f}}{4}
\frac{\partial^2 ({\rm tr} M^{-1})}{\partial \mu^2} \right\rangle
+ 2 \left\langle \left( \frac{N_{\rm f}}{4} \right)^2
\frac{\partial (\ln \det M)}{\partial \mu}
\frac{\partial ({\rm tr} M^{-1})}{\partial \mu} \right\rangle
\right. \nonumber \\
&& 
+ \left\langle \left( \frac{N_{\rm f}}{4} \right)^2
\frac{\partial^2 (\ln \det M)}{\partial \mu^2}
{\rm tr} M^{-1} \right\rangle
+ \left\langle \left( \frac{N_{\rm f}}{4} \right)^3 \left(
\frac{\partial (\ln \det M)}{\partial \mu} \right)^2 {\rm tr} M^{-1}
\right\rangle
\nonumber \\
&& \left.
- \left( \left\langle \frac{N_{\rm f}}{4}
\frac{\partial^2 (\ln \det M)}{\partial \mu^2} \right\rangle
+ \left\langle \left( \frac{N_{\rm f}}{4}
\frac{\partial (\ln \det M)}{\partial \mu} \right)^2 \right\rangle
\right) \left\langle \frac{N_{\rm f}}{4} {\rm tr} M^{-1} \right\rangle
\right] \nonumber \\
&& \hspace{-1cm}
\times\left[ \left\langle \frac{N_{\rm f}}{4}
\frac{\partial^2 (\ln \det M)}{\partial \mu^2} \right\rangle
+ \left\langle \left( \frac{N_{\rm f}}{4}
\frac{\partial (\ln \det M)}{\partial \mu} \right)^2 \right\rangle
\right].
\nonumber
\end{eqnarray}
As explained above, all terms involving the expectation value of an odd number
of derivations with respect to $\mu$ have been set to zero.
Evaluation of 
eqns. (\ref{eq:d3zdmu2dm}) and (\ref{eq:d5zdmu4dm}) requires the following 
expressions for the derivatives of $\mbox{tr}M^{-1}$:
\begin{eqnarray}
\label{eq:45}
&&\hspace{-2cm}\frac{\partial {\rm tr} M^{-1}}{\partial \mu}
=  - {\rm tr} \left( M^{-1} \frac{\partial M}{\partial \mu}
 M^{-1} \right) \\
&&\hspace{-2cm}\frac{\partial^2 {\rm tr} M^{-1}}{\partial \mu^2}
= - {\rm tr} \left( M^{-1} \frac{\partial^2 M}{\partial \mu^2}
 M^{-1} \right)
 + 2 {\rm tr} \left( M^{-1} \frac{\partial M}{\partial \mu}
    M^{-1} \frac{\partial M}{\partial \mu} M^{-1} \right) \\
&&\hspace{-2cm}
\frac{\partial^3 {\rm tr} M^{-1}}{\partial \mu^3}
= - {\rm tr} \left( M^{-1} \frac{\partial^3 M}{\partial \mu^3}
 M^{-1} \right)
 +3 {\rm tr} \left( M^{-1} \frac{\partial^2 M}{\partial \mu^2}
    M^{-1} \frac{\partial M}{\partial \mu} M^{-1} \right)  \\
&&\hspace{-0.8cm} +3 {\rm tr} \left( M^{-1} \frac{\partial M}{\partial \mu}
    M^{-1} \frac{\partial^2 M}{\partial \mu^2} M^{-1} \right)
-6 {\rm tr} \left( M^{-1} \frac{\partial M}{\partial \mu}
    M^{-1} \frac{\partial M}{\partial \mu} M^{-1}
    \frac{\partial M}{\partial \mu} M^{-1} \right) \nonumber \\
\label{eq:48}
&&\hspace{-2cm}
\frac{\partial^4 {\rm tr} M^{-1}}{\partial \mu^4}
= - {\rm tr} \left( M^{-1} \frac{\partial^4 M}{\partial \mu^4}
 M^{-1} \right)
 +4 {\rm tr} \left( M^{-1} \frac{\partial^3 M}{\partial \mu^3}
    M^{-1} \frac{\partial M}{\partial \mu} M^{-1} \right)  \\
&&\hspace{-0.8cm} 
+6 {\rm tr} \left( M^{-1} \frac{\partial^2 M}{\partial \mu^2}
    M^{-1} \frac{\partial^2 M}{\partial \mu^2} M^{-1} \right)
+4 {\rm tr} \left( M^{-1} \frac{\partial M}{\partial \mu}
    M^{-1} \frac{\partial^3 M}{\partial \mu^3} M^{-1} \right) \nonumber \\
&&\hspace{-0.8cm} 
 -12 {\rm tr} \left( M^{-1} \frac{\partial^2 M}{\partial \mu^2}
    M^{-1} \frac{\partial M}{\partial \mu} M^{-1}
    \frac{\partial M}{\partial \mu} M^{-1} \right) \nonumber \\
&&\hspace{-0.8cm} 
 -12 {\rm tr} \left( M^{-1} \frac{\partial M}{\partial \mu}
    M^{-1} \frac{\partial^2 M}{\partial \mu^2} M^{-1}
    \frac{\partial M}{\partial \mu} M^{-1} \right) 
\nonumber \\
&&\hspace{-0.8cm} 
-12 {\rm tr} \left( M^{-1} \frac{\partial M}{\partial \mu}
    M^{-1} \frac{\partial M}{\partial \mu} M^{-1}
    \frac{\partial^2 M}{\partial \mu^2} M^{-1} \right) \nonumber \\
&&\hspace{-0.8cm} 
 +24 {\rm tr} \left( M^{-1} \frac{\partial M}{\partial \mu}
    M^{-1} \frac{\partial M}{\partial \mu} M^{-1}
    \frac{\partial M}{\partial \mu} M^{-1}
    \frac{\partial M}{\partial \mu} M^{-1} \right) \nonumber
\end{eqnarray}

\section{The pressure of free staggered fermions}
\label{pressure_mu_series}
Expanding the quantity $p/T^4$ as discussed in
section~\ref{cut-off_dependence} one finds
\bqa
\lefteqn{\frac{p}{T^4}\Big|_{N_\tau} = N_{\rm f}\sum_{i=0}^{\infty}
\mathcal{C}_i\Big|_{\Nt}  \left(\frac{\mu}{T}\right)^i
\;=\; \frac{3}{8}N_{\rm f}\Bigg[\sum_{i=0}^{\infty}
\left(\frac{\Nt^{3-i}}{(2\pi)^3}\int_{0}^{2\pi}d^3\vec{p}\, \sum_{p_4}
  \,c_i(p)\, \right)\left(\frac{\mu}{T}\right)^i} \nn \\
&& - \frac{\Nt^4}{(2\pi)^4}\int_{0}^{2\pi}d^4p\,
  \,c_0(p)\, \Bigg] ~~.
\label{p_coeffs}
\eqa 
Here only the even expansion coefficients give non-vanishing
contributions. Introducing the abbreviation, 
\bq
D=4\sum_{\mu} f^2_{\mu}(p),
\eq
with $f_\mu(p)$ as given in (\ref{propp}), the even 
expansion coefficients for the standard action are given by:
\bqa
c_0 &=& \ln(D) \\
c_2 &=& \frac{1}{4D^2}\big(1 - 4D\cos(2p_4) - \cos(4p_4)\big)\\
c_4 &=& \frac{1}{96D^4}\big(-9 + 8D^2 - 8D(-3 + 4D^2)\cos(2p_4) \nn \\
    & & + (12 - 56D^2)\cos(4p_4) - 24D\cos(6p_4) - 3\cos(8p_4)\big)\\
c_6 &=& \frac{1}{2880D^6}\big(150 - 180D^2 + 32D^4 -
    8D(45 - 60D^2 \nn \\
    & & + 16D^4)\cos(2p_4) + (-225 + 960D^2 - 992D^4)\cos(4p_4) \nn \\
    & & + 540D\cos(6p_4) - 1440D^3\cos(6p_4) + 90\cos(8p_4) \nn \\
    & & - 780D^2\cos(8p_4) - 180D\cos(10p_4) - 15\cos(12p_4)\big),
\eqa
For the Naik action we introduce an additional function,
\bq
g_4(p)\equiv \left. -i\frac{d f_4(\vec{p},p_4-i\mu)}{d \mu}\right|_{\mu=0}
=-\frac{9}{16}\cos(p_4)+\frac{3}{48}\cos(3p_4),
\eq
The even expansion coefficients can then be written as:
\bqa
c_0 &=& \ln(D)\\
c_2 &=& \frac{-2}{3D^2} \big( -6Df_4^2(p) + 6Dg_4^2(p) -
        48f_4^2(p)g_4^2(p)  \nn \\
    & & + Df_4(p)\sin (3p_4) \big) \\
c_4 &=& \frac{1}{36D^4}\Big(48\big( -768f_4^4(p)g_4^4(p) +
       D^3\big( f_4^2(p) - g_4^2(p) \big)  \nn \\
    & & - 192Df_4^2(p)g_4^2(p) \big( f_4^2(p) - g_4^2(p) \big)  +
        D^2\big( -6f_4^4(p) \nn \\
    & & + 44f_4^2(p)g_4^2(p) - 6g_4^4(p) \big)  \big)
        - 24D^2\big( D - 8f_4^2(p) \big) g_4(p) \cos(3p_4) \nn \\
    & & - 32D{f_4(p)}\big( D^2 - 3Df_4^2(p) + 9Dg_4^2(p) - 48f_4^2(p)g_4^2(p) \big)
        \sin(3p_4) \nn  \\
    & & + D^2\big( D - 8f_4^2(p) \big) \sin^2(3p_4)\Big) \\
c_6 &=& \frac{1}{1620D^6}\Big(-720D^2g_4(p)\big( D^3
        + 768f_4^4(p)g_4^2(p) + D^2\big( -26f_4^2(p) \nn \\
    & & + 6g_4^2(p) \big) + 96D\big( f_4^4(p)
        - 2f_4^2(p)g_4^2(p)\big)\big) \cos(3p_4) \nn \\
    & & - 45D^4\big( D - 8f_4^2(p) \big)\cos^2(3p_4)
        - 96Df_4(p)\big( 8D^4 + 46080f_4^4(p)g_4^4(p) \nn \\
    & & - 75D^3\big( f_4^2(p) - 3g_4^2(p) \big)
        + 60D^2\big( 3f_4^4(p) - 50f_4^2(p)g_4^2(p) + 15g_4^4(p) \big) \nn \\
    & & + 2880D\big( 3f_4^4(p)g_4^2(p) - 5f_4^2(p)g_4^4(p)\big)\big)
          \sin(3p_4) + 15D^2\big( 5D^3 \nn \\
    & & + 4608f_4^4(p) g_4^2(p) + D^2 \big( -76f_4^2(p)
        + 36g_4^2(p) \big) + 192D\big( f_4^4(p) \nn \\
    & & - 6f_4^2(p)g_4^2(p) \big)\big) \sin^2(3p_4)
        + 10D^3f_4(p)\big( 3D - 16f_4^2(p)\big) \sin^3 (3p_4) \nn \\
    & & + 72\big( 4\big( 245760{f_4(p)}^6g_4^6(p) +
          D^5\big( f_4^2(p) - g_4^2(p) \big) \nn \\
    & & + 92160Df_4^4(p)g_4^4(p) \big( f_4^2(p) - g_4^2(p)\big)
        - 2D^4\big( 15f_4^4(p) - 94f_4^2(p)g_4^2(p) \nn \\
    & & + 15g_4^4(p) \big)  + 120D^3\big( f_4^6(p) -
          23f_4^4(p)g_4^2(p) + 23f_4^2(p)g_4^4(p) - g_4^6(p) \big) \nn \\
    & & + 960D^2\big( 9f_4^6(p)g_4^2(p) - 34f_4^4(p)g_4^4(p) +
          9f_4^2(p)g_4^6(p) \big)\big) \nn \\
    & & - 5D^3f_4(p) \big( 3D - 16f_4^2(p) \big) g_4(p)
        \sin (6p_4) \big) \Big)
\eqa
To simplify the expressions for the p4 action we define the
expansion coefficients recursively and thus also list the odd expansion
coefficients. However, after integration over the momenta also in this
case only even powers of $\mu/T$ contribute to the expansion of the
pressure. Introducing further abbreviations,
\bq
S_\mu=\sum_{\nu\ne\mu}\sin^2(p_\nu)
\qquad \mbox{and}\qquad
\bar{c}_k=-ic_k,
\eq
the expansion coefficients for the p4 action can be written as 
\bqa
c_0 &=& \ln(D) \\
c_1 &=& -\frac{i}{6D}\big(-S_1 - S_2 - S_3 + 6S_4^2 + S_1\cos(2p_1) 
               + S_2\cos(2p_2)  \nn \\
           & & + S_3\cos(2p_3)\big) \sin(2p_4)\\
c_2 &=& \frac{1}{18D}\big(9D\bar{c}_1^2 + 6\big(3S^2_4
                         - S_1\sin^2(p_1)
                         - S_2\sin^2(p_2) \nn \\
           & &           - S_3\sin^2(p_3)\big)\sin^2(p_4)
                  - 2\cos^2(p_4)\big(9S^2_4 \nn \\
           & &   + \sin^2(p_1) \big(-3S_1 + \sin^2(p_4)\big)
                 + \sin^2(p_2)\big(-3S_2 + \sin^2(p_4)\big) \nn \\
           & &   + \sin^2(p_3)\big(-3S_3 + \sin^2(p_4)\big)\big)\big) \\
c_3 &=& \frac{i}{18D} \big(3D\bar{c}_1\big(\bar{c}_1^2 - 6c_2\big) -
                  \big(-3 + \cos(2p_1) + \cos(2p_2) \nn \\
           & & + \cos(2p_3)\big)\cos^3(p_4)\sin(p_4) -
                 2\cos(p_4)\sin(p_4)\big(12S_4^2  \nn \\
           & & + \sin^2(p_1)\big(-4S_1 + \sin^2(p_4)\big)
               + \sin^2(p_2)\big(-4S_2 + \sin^2(p_4)\big) \nn \\
           & & + \sin^2(p_3)\big(-4S_3 + \sin^2(p_4)\big)\big)\big) \\
c_4 &=& \frac{1}{216D}\big(6\cos^4(p_4)\big(\sin^2(p_1) + \sin^2(p_2) +
               \sin^2(p_3)\big) \nn \\
           & & - 3\big(3D\big(\bar{c}_1^4 - 12\bar{c}_1^2c_2 +
               12c_2^2 - 24\bar{c}_1\bar{c}_3\big) +
               8\big(-3S_4^2 + S_1\sin^2(p_1) \nn \\
           & & + S_2\sin^2(p_2) + S_3\sin^2(p_3)\big)\sin^2(p_4) +
               \big(-3 + \cos(2p_1) \nn \\
           & & + \cos(2p_2) + \cos(2p_3)\big)\sin^4(p_4))
               - 4\cos^2(p_4)\big(18S_4^2 \nn \\
           & & + \sin^2(p_1)\big(-6S_1 + 11\sin^2(p_1)\big)
               + \sin^2(p_2)\big(-6S_2 + 11\sin^2(p_2)\big) \nn \\
           & & + \sin^2(p_3)\big(-6S_3 + 11\sin^2(p_3)\big)\big)\big) \\
c_5 &=& -\frac{i}{360D}\big(3D\big(\bar{c}_1^5 - 20c_1^3c_2
               - 60\bar{c}_1^2\bar{c}_3 + 120{c}_2\bar{c}_3
               + 60\bar{c}_1\big(c_2^2 + 2c_4\big)\big) \nn \\
           & & + 20\big(-3 + \cos(2p_1) + \cos(2p_2) + \cos(2p_3)\big)\cos^3(p_4)
               \sin(p_4) \nn \\
           & & + 8\cos(p_4)\sin(p_4) \big(12S_4^2 +
               \sin^2(p_1)\big(-4S_1 + 5\sin^2(p_4)\big) \nn \\
           & & + \sin^2(p_2)\big(-4S_2 + 5\sin^2(p_4)\big) +
               \sin^2(p_3)\big(-4S_3 + 5\sin^2(p_4)\big)\big)\big) \\
c_6 &=& -\frac{1}{6480D}\big(6 - 9D\bar{c}_1^6 + 270D\bar{c}_1^4c_2 -
               1620D\bar{c}_1^2c_2^2 + 1080Dc_2^3 \nn \\
           & & + 1080D\bar{c}_1^3\bar{c}_3 - 6480D\bar{c}_1c_2\bar{c}_3 -
               3240D\bar{c}_3^2 - 3240D\bar{c}_1^2c_4 \nn \\
           & & + 6480Dc_2c_4 - 6480D\bar{c}_1\bar{c}_5 - 2\cos(2p_1)
               - 2\cos(2p_2) - 2\cos(2p_3) \nn \\
           & & +  31\cos(2(p_1 - 2p_4)) + 31\cos(2(p_2 - 2p_4))
               +  31\cos(2(p_3 - 2p_4)) \nn \\
           & & + 24S_1\cos(2(p_1 - p_4)) +  24S_2\cos(2(p_2 - p_4))
               + 24S_3\cos(2(p_3 - p_4)) \nn \\
           & & - 48S_1\cos(2p_4) -  48S_2\cos(2p_4) -
               48S_3\cos(2p_4) + 288S_4^2\cos(2p_4) \nn \\
           & & - 186\cos(4p_4) + 24S_1\cos(2(p_1 + p_4)) + 24S_2\cos(2(p_2 +
           p_4)) \nn \\
           & & +  24S_3\cos(2(p_3 + p_4)) + 31\cos(2(p_1 + 2p_4)) \nn \\
           & & +  31\cos(2(p_2 + 2p_4)) + 31\cos(2(p_3 + 2p_4))\big).
\eqa
Note that we have defined here the coefficients $c_i$ without $\Nt$ factors, 
which can be found in front of the integrals in (\ref{p_coeffs}).
\end{appendix}

\begin{thebibliography}{99}
\baselineskip 10pt
%
%
\bibitem{QCDTARO}
S. Choe {\it et al.} [QCDTARO collaboration], Phys. Rev. {\bf D65} (2002)
054501.
%
\bibitem{FK}
Z. Fodor and S. Katz, 
JHEP {\bf 0203} (2002) 014.
%
\bibitem{us}
C.R. Allton, S. Ejiri, S.J. Hands, O. Kaczmarek, F. Karsch, E. Laermann, 
Ch. Schmidt and L. Scorzato, Phys. Rev. {\bf D66} (2002) 074507.
%
\bibitem{dFP}
P. de Forcrand and O. Philipsen, Nucl. Phys. {\bf B642} (2002) 290.
%
\bibitem{B-MMRS}
P. Braun-Munzinger, D. Magestro, K. Redlich and J. Stachel, Phys. Lett.
{\bf B518} (2001) 41.
%
\bibitem{FK2}
Z. Fodor and S.D. Katz, Phys. Lett. {\bf B534} (2002) 87.
%
\bibitem{FKS}
Z. Fodor, S.D. Katz and K.K. Szab\'o, {\tt hep-lat/0208078}.
%
\bibitem{Gottlieb}
S. Gottlieb, W. Liu, D. Toussaint, R.L. Renken and R.L. Sugar,
Phys. Rev. {\bf D38} (1988) 2888.
%
\bibitem{GG}
R.V. Gavai and S. Gupta, {\tt hep-lat/0303013}.
%
\bibitem{dEL}
M. D'Elia and M.-P. Lombardo, Phys. Rev. {\bf D67} (2003) 014505.
%
\bibitem{Laine} A. Hart, M. Laine and O. Philipsen, Phys. Lett. {\bf
B505} (2001) 141.
%
\bibitem{Christian}
Ch. Schmidt, C.R. Allton, S. Ejiri, S.J. Hands, O. Kaczmarek, F. Karsch and E.
Laermann, Nucl. Phys. B (Proc. Suppl.) {\bf 119} (2003) 517.
%
\bibitem{Naik}
S. Naik, Nucl. Phys. {\bf B316} (1989) 238.
%
\bibitem{Heller}
U.M.~Heller, F.~Karsch and B.~Sturm,
Phys. Rev. {\bf D60} (1999) 114502.
%
\bibitem{KLP}
F. Karsch, E. Laermann and A. Peikert, Phys. Lett. {\bf B478} (2000) 447.
%
\bibitem{CPPACS} 
A. Ali Khan {\it et al.}  [CP-PACS collaboration],
Phys.  Rev.  {\bf D64} (2001) 074510.
%
\bibitem{Liu}
Y. Liu, H. Matsufuru, O. Miyamura, A. Nakamura and T. Takaishi,
Soryushiron Kenkyu (Kyoto) {\bf 105} (2002) D77.
%
\bibitem{Chris}
C.R. Allton, {\tt hep-lat/9610016} and Nucl. Phys. B (Proc. Suppl.) 
{\bf53} (1997) 867.
%
\bibitem{Boyd}
G. Boyd, J. Engels, F. Karsch, E. Laermann, C. Legeland, M. Lutgemeier and
B. Petersson,
Nucl. Phys. {\bf B469} (1996) 419.
%
\bibitem{roberge}
A. Roberge and N. Weiss, Nucl. Phys. {\bf B275} (1986) 734.
%
\bibitem{gavai}
R.V. Gavai and S. Gupta, Phys. Rev. {\bf D65} (2002) 094515.
\bibitem{rebhan}
J.-P. Blaizot, E. Iancu and A. Rebhan, Phys. Lett. {\bf B523} (2001)
143.
\bibitem{blaizot}
J.-P. Blaizot, E. Iancu and A. Rebhan, Eur.\ Phys.\ J.\ C {\bf 27}
(2003) 433.
\bibitem{milc}
C.~Bernard {\it et al.}  [MILC Collaboration], {\tt hep-lat/0209079}.
\bibitem{AHMJK}
M. Asakawa, U. Heinz and B. M\"uller,
Phys. Rev. Lett. {\bf 85} (2000) 2072;\\
S. Jeon and V. Koch, 
Phys. Rev. Lett.  {\bf 85} (2000) 2076;
{\tt hep-ph/0304012}.
%
\bibitem{HS}
Y. Hatta and M.A. Stephanov, {\tt hep-ph/0302002}.
\end{thebibliography}
\end{document}